\documentclass{aa} 
\newcommand{\HI}{H\,{\scriptsize I}\:}

\usepackage{graphicx}
\usepackage{txfonts}
\usepackage{amsmath}
\usepackage{hyperref}
\hypersetup{draft} %% uncomment: resolves problem of not compiling file
\begin{document}

   %\title{Hydrodynamics of giant planet formation}
   \title{Formation of the Musca filament: Evidence for asymmetries in the accretion flow due to a cloud-cloud collision}

   %\subtitle{I. Overviewing the $\kappa$-mechanism}

   \author{L. Bonne
          \inst{1}
          %\fnmsep\thanks{Just to show the usage
          %of the elements in the author field}
          \and
          S. Bontemps \inst{1}
          \and
          N. Schneider \inst{2}
          \and
          S. D. Clarke \inst{2}
          \and
          D. Arzoumanian \inst{3}
          \and
          Y. Fukui \inst{4}
          \and
          K. Tachihara \inst{4}
          \and
          T. Csengeri \inst{5,1}
          \and
          R. Guesten \inst{5}
          \and
          A. Ohama \inst{4}
          \and
          R. Okamoto \inst{4}
          \and
          R. Simon \inst{2}
          \and
          H. Yahia \inst{6}
          \and
          H. Yamamoto \inst{4}
          }

   \institute{Laboratoire d'Astrophysique de Bordeaux, Universit\'e de Bordeaux, CNRS, B18N,
              all\'ee Geoffrey Saint-Hilaire, 33615 Pessac, France\\
              \email{lars.bonne@u-bordeaux.fr}
         \and
              I. Physikalisches Institut, Universit\"at zu K\"oln, Z\"ulpicher Str. 77, 50937 K\"oln, Germany  
              \and
              Instituto de Astrofisica e Ciencias do Espaco, Universidade do Porto, CAUP, Rua das Estrelas, PT4150-762 Porto, Portugal
              \and
              Department of Physics, Nagoya University, Chikusa-ku, Nagoya 464-8602, Japan  
              \and
              Max-Planck-Institut f\"ur Radioastronomie, Auf dem H\"ugel 69, 53121 Bonn, Germany
              \and
              INRIA Bordeaux Sud-Ouest, 33405 Talence, France
         %\and
          %   I. Physikalisches Institut, Universit\"at zu K\"oln, Z\"ulpicher Str. 77, 50937 K\"oln, Germany\\
             %\email{c.ptolemy@hipparch.uheaven.space}
             %\thanks{The university of heaven temporarily does not
             %        accept e-mails}
             }

   \date{Received \today; accepted --}

% \abstract{}{}{}{}{} 
% 5 {} token are mandatory
 
  \abstract
  % context heading (optional)
  %{} leave it empty if necessary  
   {Dense molecular filaments are ubiquituous in the interstellar medium, yet their internal physical conditions and the role of gravity, turbulence, the magnetic field, radiation, and the ambient cloud during their evolution remain debated.}
  % aims heading (mandatory)
   {We study the kinematics and physical conditions in the Musca filament, the ambient cloud, and the Chamaeleon-Musca complex to constrain the physics of filament formation.}
  % methods heading (mandatory)
   {We produced CO(2-1) isotopologue maps with the APEX telescope that cut through the Musca filament. We further study a NANTEN2 $^{12}$CO(1-0) map of the full Musca cloud, \HI emission of the Chamaeleon-Musca complex, a Planck polarisation map, line radiative tranfer models, GAIA data, and synthetic observations from filament formation simulations.}
  % results heading (mandatory)
   {%Modelling the CO emission of the Musca filament indicates that the crest is a cold ($\sim$ 10 K) and dense (n$_{H_{2}} \sim$ 10$^{4}$ cm$^{-3}$) cylindrical filament, with directly next to it independent dense gas (n$_{H_{2}} \sim$ 3$\cdot$10$^{3}$ cm$^{-3}$) with a temperature of $\sim$ 15 K, the so-called strands.
   The Musca cloud, with a size of $\sim$ 3-6 pc, contains multiple velocity components. Radiative transfer modelling of the CO emission indicates that the Musca filament consists of a cold ($\sim$10 K), dense (n$_{\rm H_2}\sim$10$^4$ cm$^{-3}$) crest, which is best described with a cylindrical geometry. Connected to the crest, a separate gas component at T$\sim$15 K and n$_{\rm H_2}\sim$10$^3$ cm$^{-3}$ is found, the so-called strands.
   The velocity-coherent filament crest has an organised transverse velocity gradient that is linked to the kinematics of the nearby ambient cloud. This velocity gradient has an angle $\ge$ 30$^{\circ}$ with respect to the local magnetic field orientation derived from Planck, and the magnitude of the velocity gradient is similar to the transonic linewidth of the filament crest. Studying the large scale kinematics, we find coherence of the asymmetric kinematics from the 50 pc \HI cloud down to the Musca filament. We also report a strong [C$^{18}$O]/[$^{13}$CO] abundance drop by an order of magnitude from the filament crest to the strands over a distance $<$ 0.2 pc in a weak ambient far-ultraviolet (FUV) field.% as well as a low NH$_{3}$ abundance ([NH$_{3}$]/[H$_{2}$] $<$ 10$^{-9}$) for the filament crest.
   }
  % conclusions heading (optional), leave it empty if necessary 
   {The dense Musca filament crest is a long-lived (several crossing times), dynamic structure that can form stars in the near future because of continuous mass accretion replenishing the filament. %This mass accretion appears to be driven by a colliding \HI flow that forms the Chamaeleon-Musca complex. We propose that this colliding flow is a \HI cloud-cloud collision which leads to bending of the magnetic field around dense filaments. This bending is then responsible for the observed asymmetric accretion scenario of the Musca filament, which is for instance seen as a V-shape in the PV diagram.
   This mass accretion on the filament appears to be triggered by a \HI cloud-cloud collision, which bends the magnetic field around dense filaments. This bending of the magnetic field is then responsible for the observed asymmetric accretion scenario of the Musca filament, which is, for instance, seen as a V-shape in the position-velocity (PV) diagram.}


   \maketitle
%
%-------------------------------------------------------------------
\section{Introduction}
\begin{figure*}
\begin{center}
\includegraphics[width=0.9\hsize]{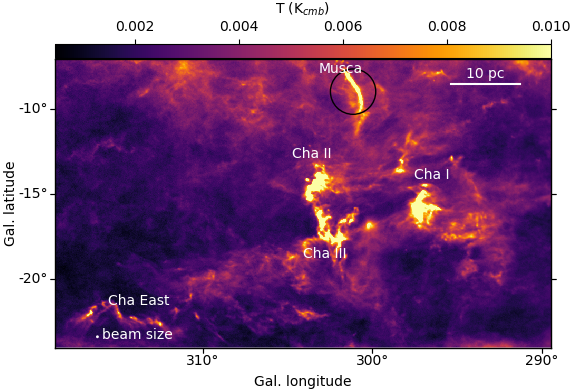}
\end{center}
\caption{Planck emission map at 353 GHz of the Chamealeon-Musca complex \citep{PlanckCollaboration2016data}. The names of the relatively dense regions in the complex are indicated in white. The black circle indicates the region where GAIA data were extracted to investigate the distance of the Musca filament and cloud. The Planck resolution at 353 GHz is 4.8$^{\prime}$, which corresponds to a physical size of 0.25 pc at a distance of 180 pc \citep{Zucker2019}.}
\label{PlanckMapChamaeleon}
\end{figure*}
The complexity of the interstellar medium (ISM) has been revealed in numerous continuum and molecular line studies, but it is only since the unprecedented far-infrared sensitivity of the \textit{\textit{Herschel} Space Telescope} \citep{Pilbratt2010} that the ubiquity of dense filamentary structures in the ISM has been revealed and their integral role in the star formation process has been established. It was shown that almost all pre-stellar and protostellar cores
are located in filaments or at a filament junction \citep[e.g.][]{Andre2010,Molinari2010,Bontemps2010,Konyves2010,Konyves2015,Arzoumanian2011,Hill2011,Schneider2012,Rygl2013,Polychroni2013,Andre2014,Marsh2016}. Spectral line observations of filamentary structures have shown that many dust continuum filaments contain several velocity-coherent sub-filaments, the so-called fibers \citep[e.g.][]{Hacar2013,Tafalla2015,Dhabal2018}. However, recently some theoretical studies have indicated that coherent structures in velocity space are not necessarily coherent in three-dimensional space \citep{ZamoraAviles2017,Clarke2018}. Massive filamentary structures in more distant regions, such as massive ridges \citep[e.g.][]{Schneider2010,Hennemann2012} and hub-filament systems \citep[e.g.][]{Myers2009,Schneider2012,Peretto2013,Peretto2014,Henshaw2017,Williams2018}, are proposed to be the dominant way to form rich clusters of stars \citep[e.g.][]{Motte2018}, and thus the bulk of star formation in our galaxy. 
It is therefore critical to unveil the precise physical processes at work to explain the formation of filaments of all types. \\\\
%It is thus critical to first understand the nature of filaments, e.g.: low-mass filaments in the nearby star-forming clouds \citep{Andre2014}, hub-filament systems \citep[e.g.][]{Myers2009,Schneider2012,Peretto2013,Williams2018}, massive ridges \citep[e.g.][]{Schneider2010,Hennemann2012} or galactic bones \citep[e.g.][]{Jackson2010,Wang2015}. Obviously, the true nature of filaments depends on the physical processes that govern the structure of the interstellar medium in general. 
While early theoretical studies described filaments as structures confined by isotropic pressure equilibrium \citep{Ostriker1964,Inutsuka1992}, simulations support the argument that filaments are a manifestation of structure development caused by the thermodynamic evolution of the ISM during molecular cloud formation. All (magneto)-hydrodynamic ISM simulations with turbulence, including and not including self-gravity and/or a magnetic field, naturally produce filaments \citep[ e.g.][]{Audit2005,Heitsch2005,Hennebelle2008,Nakamura2008,Banerjee2009,Gomez2014,Smith2014,Smith2016,Chen2014,Seifried2015,Federrath2016,DuarteCabral2017}. % but only a few studies are devoted to explaining the origin of these filamentary structure in these models. 
In these simulations, filaments are argued to originate from the collision of shocked sheets in turbulent flows \citep{Padoan2001}, by instabilities in
self-gravitating sheets \citep{Nagai1998}, or as the result of long-lived coherent flows in the turbulent ISM \citep{Hennebelle2013} which could be at least partly convergent. The magnetic field can only reinforce the presence of filaments, as it increases local axisymetries and as it may stabilise and drive turbulent flows aligned with the field \citep{Hennebelle2013}. The case of magnetised \HI colliding streams and cloud-cloud collisions \citep[e.g.][]{Ballesteros-Paredes1999,Koyama2002} is particularly interesting in this context as it naturally creates a significant level of turbulence and flows. It can also lead to bending of the magnetic field around pre-existing structures, which could then drive local convergent flows perpendicular to the dense filamentary structures \citep{Inoue2013,Inoue2018}. Such local convergent flows were indeed observed for massive dense cores by \cite{Csengeri2011}, for instance. From an observational point of view, cloud-cloud collisions at a velocity of 10-20 km s$^{-1}$ have been argued to form massive star-forming filaments \citep[][and references therein]{Fukui2019}.\\\\
%This so-called Inoue's scenario appears as the dominant process for filament formation in the cloud-cloud collisions \citep{Inoue2018, Fukui2019}.%or by a velocity flow generated by the bending of the magnetic field in a cloud collision \citep{Inoue2013,Inoue2018}. 
Any scenario to form filaments has to account for the most critical properties of observed filaments. The so-called universal \citep{Arzoumanian2011,Arzoumanian2019,Koch2015}, yet highly debated \citep[e.g.][]{Panopoulou2017,Seifried2017,Ossenkopf2019}, filament width of 0.1 pc in nearby molecular clouds is close to the sonic scale which could fit in a scenario where filaments are made of dense, post-shock gas of converging flows \citep[][]{Arzoumanian2011,Schneider2011,Federrath2016}. In the companion paper \citep{Bonne2020}, observational indications were found of warm gas from low-velocity shocks associated with mass accretion on the Musca filament. It has also led to some theoretical models considering gravitational inflow that might provide an explanation for this universal width \citep{Heitsch2013a,HennebelleAndre2013}. Indications of an inflowing mass reservoir as a result of gravity were presented in \cite{Palmeirim2013} and \cite{Shimajiri2018}. In the medium surrounding dense filaments,  striations are often found which are well aligned with the magnetic field \citep{Goldsmith2008,Palmeirim2013,AlvesDeOliveira2014,Cox2016,Malinen2016}. This has often led to the interpretation of gas streaming along the magnetic field lines. So far, there is no strong observational evidence for this and recently it was argued from a theoretical study that striations might %not be flow structures but rather the 
result from MHD waves \citep{Tritsis2016b} in large pc-scale sheets. Similarly, the so-called fibers, seen as sub-structures of filaments, are found at different velocities as if they would originate from slightly different velocity flows inside the same global convergence of flows in the cloud. Several numerical simulations also observe fibers \citep[e.g.][]{Smith2014,Smith2016,Moeckel2015,Clarke2017,ZamoraAviles2017}, while in pressure equilibrium models filaments naturally tend to fragment rapidly into cores \citep{Inutsuka1997}.\\\\
% However, theoretical models considering a filament in quasi-static pressure equilibrium with the ambient cloud also find that a FWHM near 0.1 pc is plausible \citep{Fischera2012}.\\\\
%It thus appears that the relation between the observed filaments and the surrounding ambient cloud, in the context of understanding the precise physical origin of star-forming filaments, remains a highly debated question. To address this question,
In summary, these different views on the relation between filaments
and the surrounding ambient cloud, in particular the understanding of
the physical origin of star-forming filaments, require detailed
observational studies and comparison with simulations. In this paper, we study the Musca filament which is probably at an early evolutionary state as its ambient cloud is not yet perturbed by star formation.\\
The following section introduces the Musca filament and why this filament is particulary interesting to study the relation with its ambient cloud. Section 3 describes the observations carried out with the APEX telescope and Section 4 presents the first results of the APEX and NANTEN2 observations. Section 5 provides a radiative tranfer analysis of the observational data to constrain the physical conditions in the filament. In Section 6 the implications of these results on the formation of the Musca filament and its relation with the ambient cloud are discussed. 

%\section{Multifractal analysis of the Musca \textit{Herschel} maps}
\section{The Musca filament and Chamaeleon clouds}
\begin{figure}
\includegraphics[width=\hsize]{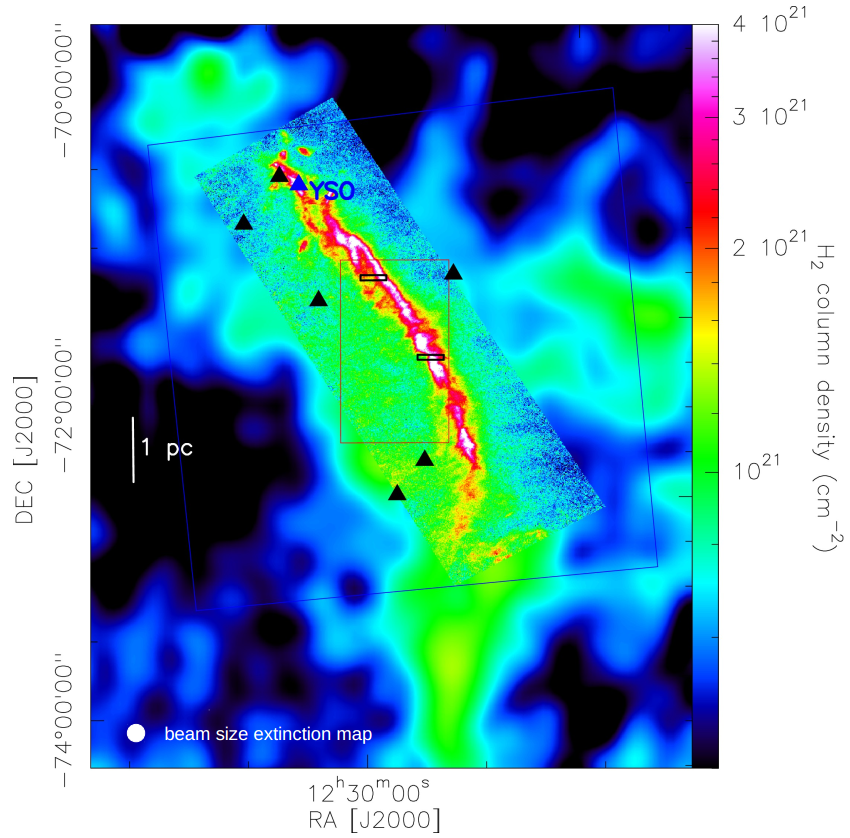}
\caption{Column density map of the Musca filament from the \textit{Herschel} Gould Belt Survey (HGBS) \citep{Andre2010,Cox2016} embedded in the large scale 2MASS extinction map of Musca, scaled to the \textit{Herschel} column density, that traces the ambient cloud. The extinction map was produced by the A$_{V}$ mapping tool in \cite{Schneider2011}. The black boxes indicate the maps made with the APEX telescope, the red box indicates the area displayed in Fig. \ref{Map250}, and the blue box indicates the area mapped with the NANTEN2 telescope. The black triangles show the locations of the stars with a distance smaller than 140 pc and a significant reddening ($>$ 0.3) in the GAIA catalogue. The blue triangle shows the location of the only young stellar object (YSO) in the Musca filament.}
\label{columnDensityMap}
\end{figure}
\begin{figure}
\includegraphics[width=\hsize]{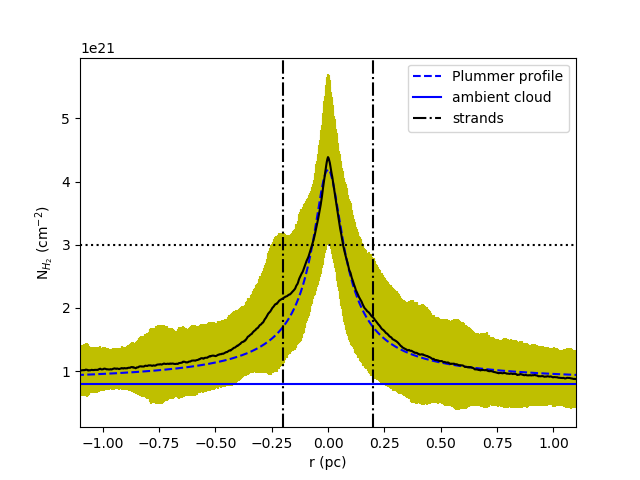}
\caption{Average column density profile of the Musca filament from the \textit{Herschel} data \citep{Cox2016}, corrected to a distance of 140 pc. The blue horizontal line indicates the column density associated with the ambient cloud (N$_{H_{2}}$ $\sim$ 0.8$\cdot$10$^{21}$ cm$^{-2}$). The dashed blue line shows the fitted Plummer profile to the column density (excluding the strands). The vertical lines indicate the location of column density excess with respect to the Plummer fit close to the filament crest, the so-called strands. The dashed horizontal black line indicates N$_{H_{2}}$ = 3$\cdot$10$^{21}$ cm$^{-2}$, which is the minimal column density used to define the filament crest.}
%\label{Map250}
\label{colDensDoris}
\end{figure}
\begin{figure}
\includegraphics[width=\hsize]{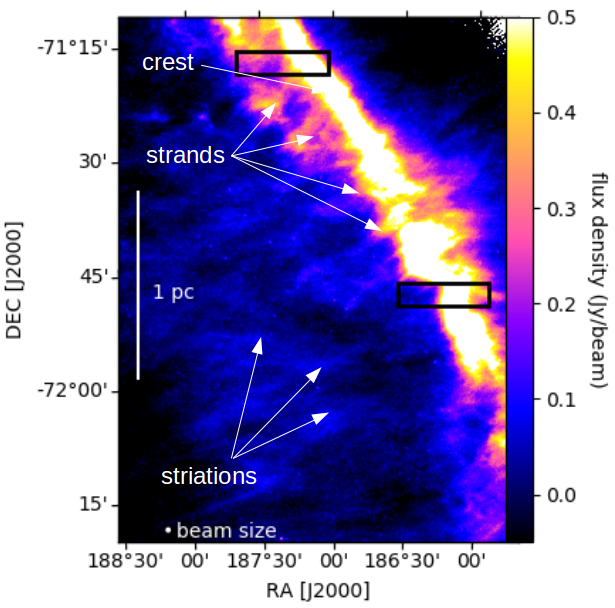}
\caption{Zoom (red box in Fig. \ref{columnDensityMap}) in the \textit{Herschel} 250 $\mu$m map of the Musca cloud \citep{Cox2016}. The hair like structures perpendicular to the Musca filament, in the strands and ambient cloud, are indicated in white. The black boxes indicate the regions covered by the APEX maps.}
\label{Map250}
\end{figure}
\subsection{The Musca filament}
We focus on the Musca filament which is seen on the sky as a 6 pc long filamentary structure in continuum, extinction and molecular lines \citep[e.g.][]{Mizuno2001,Kainulainen2009,Schneider2011,Hacar2016,Cox2016}. This filament is located in the Chamaeleon-Musca complex, see Fig. \ref{PlanckMapChamaeleon}, which is considered a single molecular complex with a size of $\sim$ 70 pc on the plane of the sky if one includes Cham East \citep[e.g.][]{Corradi1997,Mizuno2001,PlanckGrenier2015,Liszt2019}. Figure \ref{columnDensityMap} presents the column density map of the Musca filament combining 2MASS and \textit{Herschel} data which shows that this high column density filament with its ambient cloud is in relative isolation in the plane of the sky. The Musca filament hosts one protostellar core \citep{VilasBoas1994,Juvela2012,Machaieie2017}, and may contain a few prestellar cores \citep{Kainulainen2016} with an average core separation that could fit with gravitational fragmentation inside a filamentary crest. The filament is thus likely at a relatively early evolutionary stage and has indeed a relatively low line mass compared to other more active star forming filaments like B211/3 in Taurus \citep{Palmeirim2013,Cox2016,Kainulainen2016}. As there is only one protostar which is located in the far north of the filament, the cloud is still unperturbed by protostellar feedback and is thus a very interesting location to study the formation of  star forming gas and the role of the ambient cloud. The presence and structure of this ambient cloud was clearly established with high sensitivity dust continuum observations with \textit{\textit{Herschel}} and Planck \citep{Cox2016,PlanckArzoumanian2016}, and will be described in more detail below. %see Figs. \ref{PlanckMapChamaeleon}, \ref{columnDensityMap} and \ref{Map250}. With \textit{Herschel} it was shown that the ambient cloud contains a network of striations \citep{Cox2016}. 

For clarity in this paper, we distinguish four features of the Musca cloud based on the column density map and profile of dust emission, mostly following the nomenclature introduced by \citet{Cox2016}.\\
Firstly, the \underline{{\sl (filament) crest}} is the high column density spine of the large filament in the Musca cloud with N$_{H_{2}} >$ 3$\cdot10^{21}$ cm$^{-2}$ \citep{Cox2016}, see Fig. \ref{colDensDoris}. From fitting a Plummer profile to the average column density profile of the full Musca filament, working with a distance of 140 pc, this suggest a radius of 0.056 pc for the filament crest \citep{Cox2016}.\\
Secondly, \underline{{\sl strands}} were defined in \cite{Cox2016} as the immediate surrounding of the filament crest where dust column density was found to be significant (N$_{H_{2}} \sim$ 2$\cdot10^{21}$ cm$^{-2}$) and inhomogeneous. For instance, there is a strong tendency to be asymmetric with a brighter emission directly east of the filament crest, and the strands display a hair-like structure that is preferentially perpendicular to the filament crest \citep{Cox2016}, see Fig. \ref{Map250}. To go one step further we here propose that the strands could represent all column density in excess in Fig. \ref{colDensDoris} to the average Plummer profile fitted on the column density towards the crest. %are defined as the column density excess of the fitted Plummer profile that are directly next to the filament crest
These strands might then contain the most nearby ambient gas of the filament crest and may represent a mass reservoir to be collected soon by the crest. In Fig. \ref{colDensDoris} and local cuts perpendicular to the filament axis, we found that the strands typically extend up to a distance of $\sim$ 0.4 pc from the filament crest. It is not straightforward to fit a function to the strands, but from Fig. \ref{colDensDoris} it can be observed that the column density excess to the Plummer fit typically extends up to a distance of 0.4 pc. The presence of the strands leads to an asymmetric column density profile for Musca. Other asymmetric column density profiles were already observed in the Pipe Nebula \citep{Peretto2012}.\\ %Generally speaking the strands display a hair-like structure that is preferentially perpendicular to the filament crest \citep[][Fig. \ref{Map250}]{Cox2016}.\\
%$\bullet$ \underline{The {\sl filament}} corresponds to the crest + the strands, and can be defined as the large filament seen in the Musca cloud with  N$_{H_{2}} >$ 1.5$\cdot10^{21}$ cm$^{-2}$. This $'$Musca filament$'$ covers a radial area up to 0.4 pc.\\
\underline{{\sl The ambient Musca cloud}} then embeds the filament crest and strands, and has a typical average column density of N$_{H_{2}} \sim$ 0.8$\cdot$10$^{21}$ cm$^{-2}$ within a typical distance of 3 pc from the filament crest, see Figs. \ref{columnDensityMap} and \ref{colDensDoris}.\\
Lastly, \underline{{\sl Striations}} are %low column density (N$_{H_{2}} <$ 1.5$\cdot10^{21}$ cm$^{-2}$)
filamentary structures in the parsec scale ambient cloud and sub-parsec scale strands of the Musca filament, see Fig. \ref{Map250}. %The striations are continuous features through the ambient cloud and the strands, suggesting both structures are connected. In contrast, it is less clear whether the striations are present inside the crest as the crest is too narrow to firmly separate striations inside it from possible foreground emission from the Musca cloud and strands. %and generally appear perpendicular to the filament crest.
%\\
%We will thus use the same definition for strands and striations as in \cite{Cox2016}. 
The striations are well aligned with the magnetic field on the plane of the sky \citep{Cox2016}. This magnetic field is roughly perpendicular to the Musca filament crest, as is also found for other dense filamentary structures \citep{Planck2016Soler,PlanckArzoumanian2016} and even at the centre of massive ridges like DR21 \citep{Vallee2006,Schneider2010}. Using C$^{18}$O observations it was proposed that the filament crest is a velocity-coherent structure \citep{Hacar2016}, or a single fiber following the nomenclature of \cite{Hacar2013}, unlike well studied filaments such as B211/3 in Taurus and Serpens south \citep{Hacar2013,Dhabal2018}. This makes the Musca filament relatively simple in velocity space.\\
More recently, it was questioned whether the Musca filament can be described as truely cylindrical (in constrast to a sheet seen edge-on) by proposing that the striations pattern is reflecting  magnetohydrodynamic vibrations which would require a large, pc-scale region of emission in contradiction to the 0.1 pc width of the crest. It would then suggest that the Musca cloud is a sheet that is seen edge-on and thus only appears filamentary due to projection \citep{Tritsis2018}. We use our new observations and an updated view of the global structure of the cloud to re-discuss this important issue. Meanwhile we continue to use the term $^{\prime}$filament$^{\prime}$, at least for the crest, throughout the paper. %since this discussion is not settled and because the analysis of our observations presented here favor the scenario where the Musca filament crest is a cylindrical filament.

\subsection{Distance of the Musca filament}
\label{sec: distMuscaSec}
\begin{figure}
\includegraphics[width=\hsize]{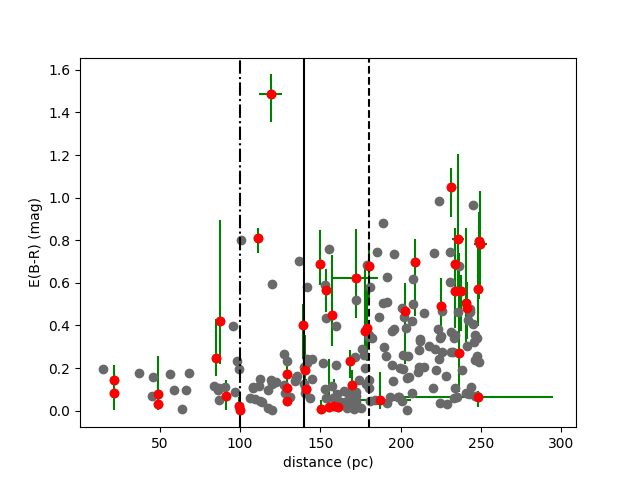}
\caption{Reddening obtained with the GAIA telescope as a function of distance for stars in the region of the Musca cloud. The coloured points are observed at locations where the Planck 353 GHz emission is above the threshold that was used to define the area covered by the Musca cloud, see App. \ref{app: distanceAppendix}. This shows a general increase of the reddening at a distance of 140 pc, shown with the full vertical line, but there also appears to be some reddening starting at 90-100 pc, indicated with the dashdot vertical line. The dashed vertical line indicates the distance of the Chamaeleon I, II and III molecular clouds at 183 pc from \citet{Zucker2019}. %\textbf{Bottom}: The column density map of the Musca filament. The blue circle indicates the region on the sky from which stars were selected in the GAIA DR2 catalogue to estimate the distance of the Musca filament. The red triangles show the locations of the stars with a distance smaller than 140 pc that have a reddening higher than 0.3 in the GAIA catalogue.
}
\label{distMusca}
\end{figure}

In earlier studies, the distance of the Musca cloud was generally estimated to be 140-150 pc \citep[e.g.][]{Franco1991,Knude1998}. Studying the reddening of stars in the GAIA DR2 data release \citep{GaiaDR2,Andrae2018} that are close to the Musca filament, with a method similar to \cite{Yan2019}, shows indeed a noteworthy reddening increase at a distance of 140-150 pc, see Fig. \ref{distMusca}. The region in the plane of the sky studied with GAIA data is displayed in Fig. \ref{PlanckMapChamaeleon}. %This tends to confirm the earlier estimated distance of 140-150 pc for the Musca filament. 
For more information on the determination of the distance, see App. \ref{app: distanceAppendix}. It can however be noted in Fig. \ref{distMusca} that there is already reddening for a couple of stars starting at 100 pc towards the Musca cloud. One should note that there are significant uncertainties on the reddening in the GAIA catalogue, but this behaviour can also be noted for two stars in \cite{Franco1991}. In the GAIA catalogue, the location of the stars with significant reddening at a distance $<$ 140 pc is spread over the full Musca cloud, see Fig. \ref{columnDensityMap}. This points to the fact that there might be some extended nearby gas at a distance of $\sim$ 100 pc towards the Musca region. %Combining Fermi LAT, \HI and Planck observations, a siginficant amount CO-dark H$_{2}$ gas was found towards the Chameleon-Musca complex \citep{PlanckGrenier2015}. Some of this CO-dark gas of the Chamaeleon-Musca complex could well be located significantly closer than the filament, 
Since the Chamaeleon-Musca complex has a size of $\sim$ 70 pc in the plane of the sky, it is not unlikely that the cloud has a similar size along the line of sight. This could possibly lead to extinction starting at 100 pc for some stars. However, the clearest jump in reddening happens around 140 pc, which is consistent with several earlier studies \citep{Franco1991,Corradi1997,Whittet1997,Knude1998} of the high column density Musca cloud. We thus assume a distance of 140 pc for the Musca filament in this paper. Note that this distance is smaller than the distance of 183 $\pm$ 3 $\pm$ 9 pc derived for the dense Chamaeleon clouds (Cha I, Cha II and Cha III) in \cite{Zucker2019}. Inspecting the reddening as a function of distance with the GAIA data, we also found a distance of 180-190 pc for these dense Chamaeleon clouds, see App. \ref{app: distanceAppendix}. This would indicate that Musca has a slightly more nearby location than the Chamaeleon clouds. 

\section{Observations}

\subsection{APEX: PI230 observations}
In September 2018, we performed observations with the PI230 receiver which is installed on the APEX telescope \citep{Guesten2006}. The two bands of PI230, each with $\sim$ 8 GHz bandwidth, were tuned to a frequency to cover 213.7-221.4 GHz and 229.5-237.2 GHz such that the $^{12}$CO(2-1), $^{13}$CO(2-1), C$^{18}$O(2-1) and SiO(5-4) lines, respectively at 230.538, 220.399, 219.560 and 217.105 GHz, were observed simultaneously. The observations with PI230 were performed in the on-the-fly mode, creating two maps in all these lines with a size of 600$^{\prime\prime} \times$ 100$^{\prime\prime}$. These two maps cover the Musca filament crest and its nearby strands in two different regions, see Fig \ref{columnDensityMap}. The northern map of the two is located towards a region that has a filamentary shape and the strands to the east, while the southern map is located in a region that shows signs of fragmentation with strands to the west. The OFF position used for the observations is located at $\alpha_{\text{(2000)}}$ = 12$^{h}$41$^{m}$38$^{s}$; $\delta_{\text{(2000)}}$ = -71$^{o}$11$^{\prime}$00$^{\prime\prime}$, identical to the OFF position used for the observations presented in \cite{Hacar2016}. This OFF position was checked using another OFF position that is located further away from the Musca filament, at $\alpha_{\text{(2000)}}$ =  12$^{h}$42$^{m}$21$^{s}$; $\delta_{\text{(2000)}}$ = -72$^{o}$27$^{\prime}$31$^{\prime\prime}$, which  was selected based on Planck maps. We found no contamination for any of the lines, for example, for $^{12}$CO(2-1) the baseline rms was 2.8$\cdot$10$^{-2}$ K.\\
The spectral resolution of these observations is $\sim$ 0.08 km s$^{-1}$, which resolves the velocity-coherent filament crest component with $\sigma \sim$ 0.15 km s$^{-1}$ presented in \cite{Hacar2016}. The spatial resolution of the observations is $\sim$ 28$^{\prime\prime}$, and inside a velocity interval of 0.1 km s$^{-1}$, the typical rms is $\sim$ 0.07 K. The main beam efficiency is $\eta_{\text{mb}}$ = 0.68 \footnote{\url{http://www.apex-telescope.org/telescope/efficiency/}}, and the forward efficiency used is 0.95. The data reduction was performed using the CLASS software\footnote{\url{http://www.iram.fr/IRAMFR/GILDAS}}.
\begin{figure*}
\begin{center}
\includegraphics[width=0.85\hsize]{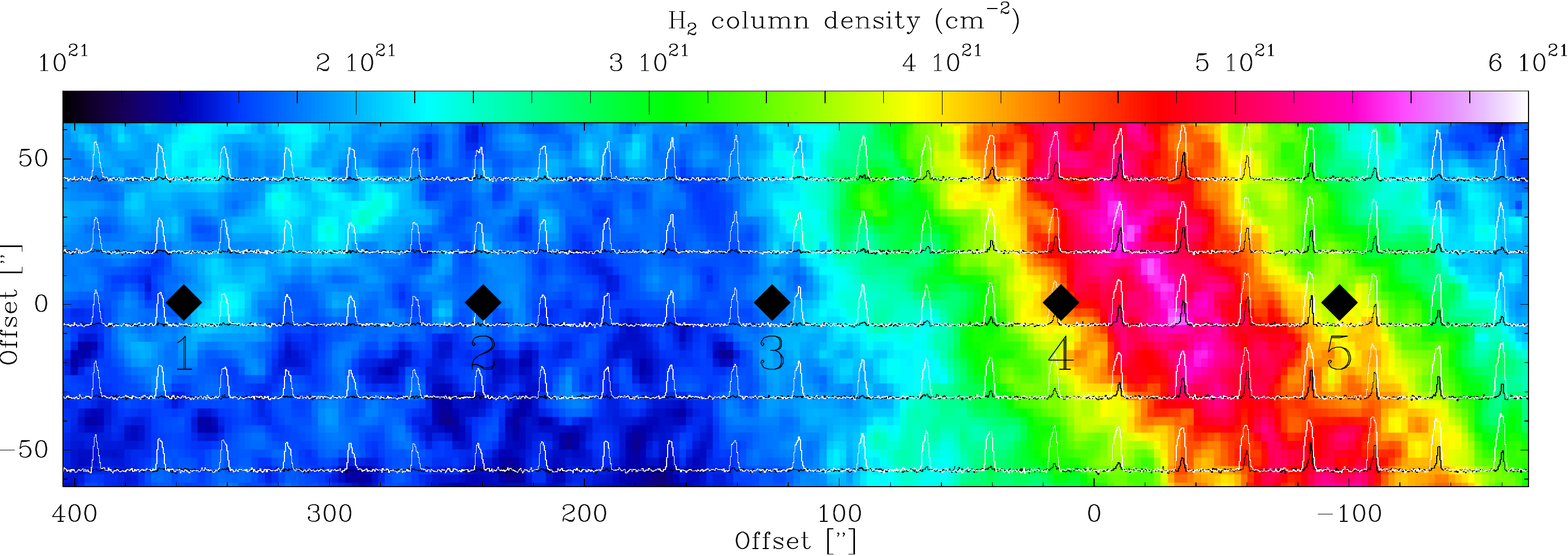}
\includegraphics[width=0.43\hsize]{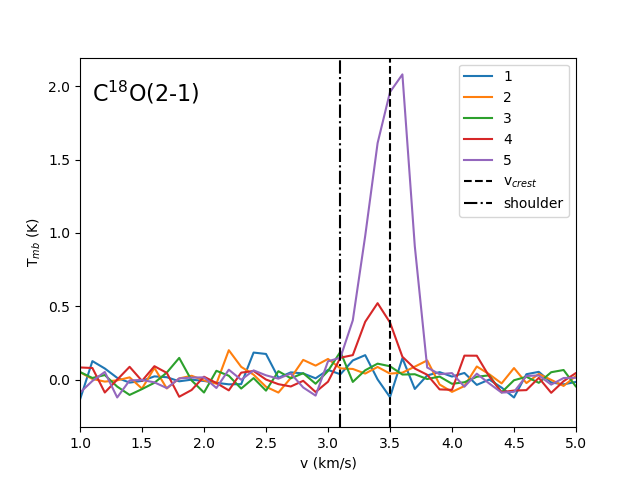}
\includegraphics[width=0.43\hsize]{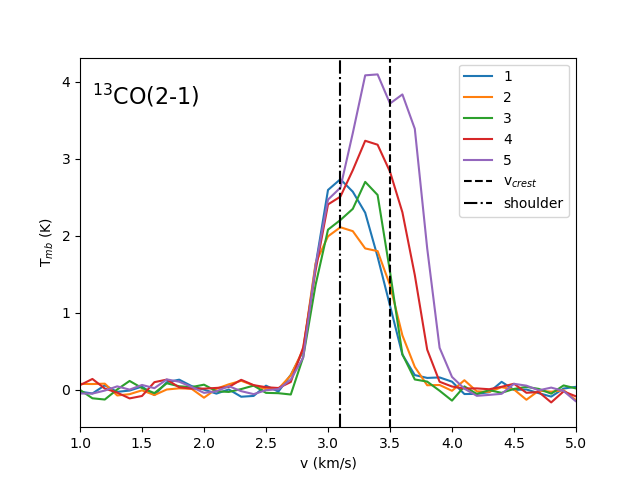}
\end{center}
\caption{\textbf{Top}: $^{13}$CO(2-1) (in white) and C$^{18}$O(2-1) (in black) spectra overlayed on the \textit{\textit{Herschel}} column density map of Musca \citep{Cox2016} for the northern position. The offset is centred on: $\alpha_{\text{(2000)}}$ = 12$^{h}$28$^{m}$58$^{s}$ and $\delta_{\text{(2000)}}$ = -71$^{o}$16$^{\prime}$55$^{\prime\prime}$, and 100$^{\prime\prime}$ corresponds to $\sim$ 0.07 pc at a distance of 140 pc. C$^{18}$O emission is only detected at the crest (green-yellow-red), while $^{13}$CO can be used to trace the strands (blue) as well. \textbf{Bottom right}: Several $^{13}$CO(2-1) spectra are displayed that were extracted at the indicated positions in the map above. This shows that $^{13}$CO has two components towards the filament crest. \textbf{Bottom left}: The same for C$^{18}$O(2-1), demonstrating it is only detected towards the filament crest.}
\label{spectraMapNorth}
\end{figure*}
\begin{figure*}
\begin{center}
\includegraphics[width=0.85\hsize]{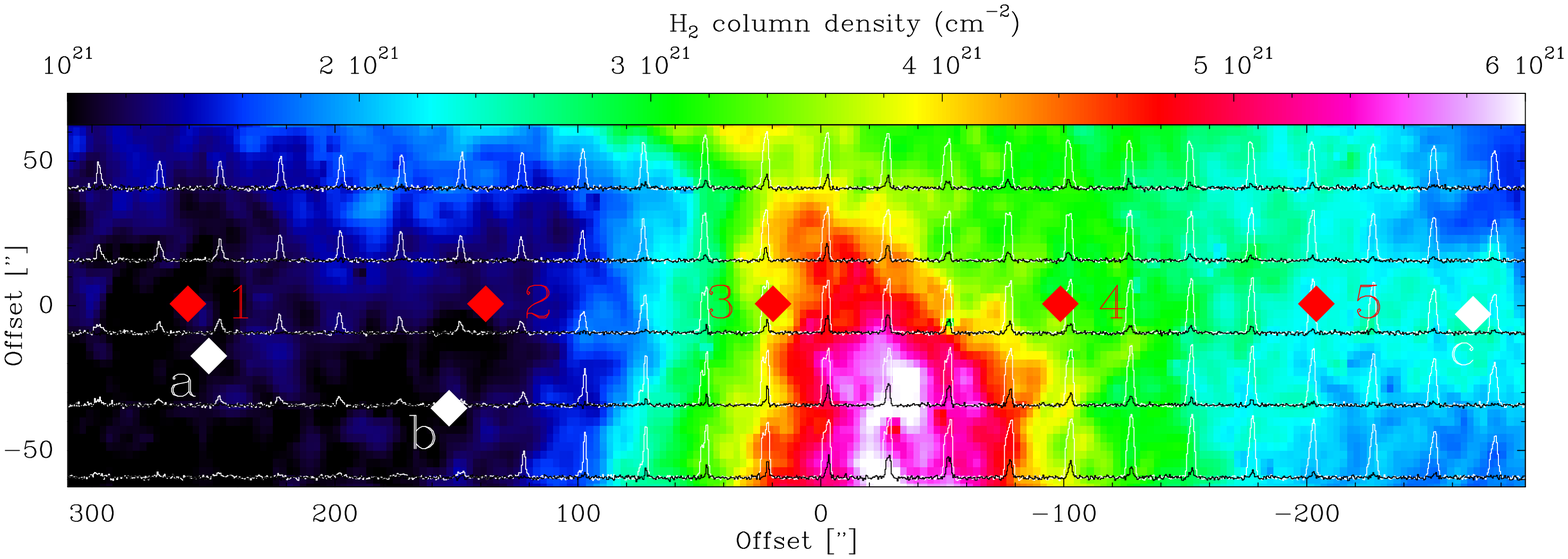}
\includegraphics[width=0.43\hsize]{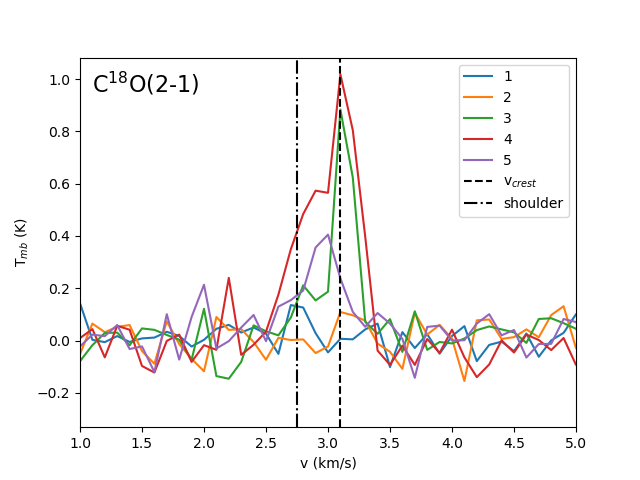}
\includegraphics[width=0.43\hsize]{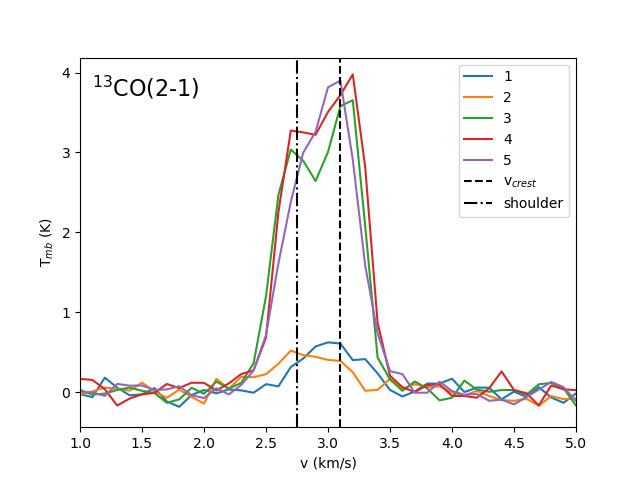}
\includegraphics[width=\hsize]{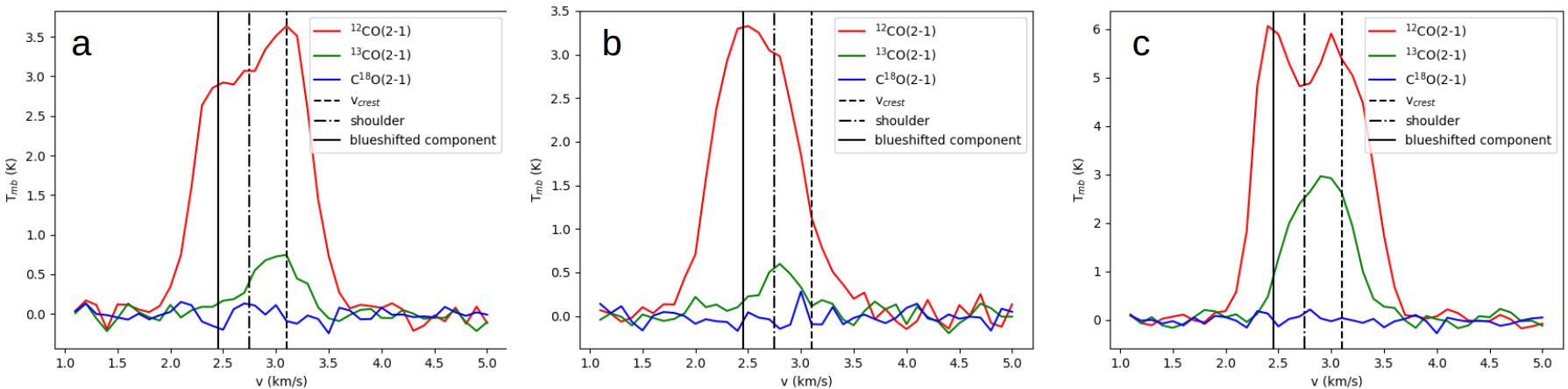}
\end{center}
\caption{\textbf{Top \& middle}: Same as Fig. \ref{spectraMapNorth}, but for the southern position, where the offset is centered on: $\alpha_{\text{(2000)}}$ = 12$^{h}$24$^{m}$46$^{s}$ and $\delta_{\text{(2000)}}$ = -71$^{o}$47$^{\prime}$20$^{\prime\prime}$. It can be noted that at the location where there are no strands (N$_{H_{2}} <$ 1.5$\cdot$10$^{21}$ cm$^{-2}$) that the $^{13}$CO emission also disappears. \textbf{Bottom}: The $^{12}$CO(2-1), $^{13}$CO(2-1) and C$^{18}$O(2-1) emission at the indicated white positions (a, b, c) in the map above with the same beam size (28$^{\prime\prime}$) and spectral resolution (0.1 km s$^{-1}$). This indicates that in $^{12}$CO(2-1) there is an extra blueshifted component that is barely detected in $^{13}$CO(2-1). It also shows that C$^{18}$O(2-1) is not detected away from the filament crest with this data.}
\label{spectraMapSouth}
\end{figure*}

\subsection{APEX: FLASH$^{+}$ observations}
With FLASH$^{+}$ \citep{Klein2014} on the APEX telescope, the $^{12}$CO(3-2), $^{12}$CO(4-3) and $^{13}$CO(3-2) lines at 345.796 GHz, 461.041 GHz and 330.588 GHz, respectively were observed simultaneously. FLASH$^{+}$ consists out of two receivers that can observe simultaneously: FLASH345 and FLASH460. Both receivers have two bands with a 4 GHz bandwidth. FLASH345 can observe in the 268-374 GHz range, while FLASH460 can observe in the 374-516 GHz range. These observations were performed towards the northern and southern map. 
%However, the observations did not cover the full size of the PI230 maps since they are much more time consuming, in particular the FLASH460 observations of $^{12}$CO(4-3). 
Towards the northern area, the FLASH$^{+}$ observations make a map of 500$^{\prime\prime} \times$ 100$^{\prime\prime}$ that covers an area including the filament crest and the eastern strands, and in the southern area the FLASH$^{+}$ map only covers an area of 120$^{\prime\prime} \times$ 100$^{\prime\prime}$ which is centered on the filament crest. The observations towards the southern map were performed in a setup that did not cover the $^{13}$CO(3-2) line. As a result this line is only available for the northern map. These observations with FLASH$^{+}$ were spread over 3 observing periods: July 2017 (P100), from May to June 2018 (P101) and September 2018 (P102). All observations experienced certain complications requiring specific attention. The P100 $^{12}$CO(3-2) observations have some contamination from the used OFF position (at $\alpha_{\text{(2000)}}$ = 12$^{h}$24$^{m}$05$^{s}$; $\delta_{\text{(2000)}}$ = -71$^{o}$23$^{\prime}$45$^{\prime\prime}$). For $^{12}$CO(3-2) a correction for the contamination from the OFF position was carried out by fitting a gaussian to the contamination in the OFF position and adding it to all spectra observed with this OFF position. This gaussian has T$_{A}^{*}$ = 1.17 K, v = 3.05 km s$^{-1}$ and a FWHM = 0.418 km s$^{-1}$. For the $^{12}$CO(4-3) observations, no contamination was found in this OFF position at a baseline rms of 0.09 K within 0.1 km s$^{-1}$. In P101, both the observations with FLASH345 and the FLASH460 instruments were shifted by respectively  -230 kHz and 490 kHz (F. Wyrowski, priv. comm.). In P102, the shift for the FLASH345 observations was solved, however the FLASH460 observations still had the same shift of 490 kHz (F. Wyrowski, priv. comm.). A correction was performed for these frequency shifts. The OFF position used in P101 and P102 is at $\alpha_{\text{(2000)}}$ = 12$^{h}$25$^{m}$15$^{s}$ and $\delta_{\text{(2000)}}$ = -71$^{o}$15$^{\prime}$21$^{\prime\prime}$, which is free of $^{12}$CO(3-2) emission with a baseline rms of T$_{mb}$ = 0.04 K within 0.06 km s$^{-1}$.\\
The observations with FLASH345 have a spectral resolution of 0.033 km s$^{-1}$ and an angular resolution of $\sim$ 18$^{\prime\prime}$. The FLASH460 observations have a spectral resolution of 0.05 km s$^{-1}$ and an angular resolution of $\sim$ 14$^{\prime\prime}$. For further analysis, all the observations (both with PI230 and FLASH$^{+}$) were sampled at the same spectral resolution of 0.1 km s$^{-1}$ since this is sufficient for Musca while it reduces the rms of CO(4-3) to $\sim$ 0.2 K and $\sim$ 0.5 K for the northern and southern map, respectively. This difference in data quality is due to different weather conditions with the water vapour varying between pwv = 0.4 and pwv = 1.0. Generally speaking, the observations for the northern map were carried out under better weather conditions. The main beam efficiencies\footnote{\url{http://www.apex-telescope.org/telescope/efficiency/}} used for the FLASH345 and FLASH460 observations are $\eta_{\text{mb}}$ = 0.65 and $\eta_{\text{mb}}$ = 0.49, respectively. The results from the $^{12}$CO(3-2) and $^{12}$CO(4-3) transitions are presented in the companion paper \citep[][hereafter Paper II]{Bonne2020}, while the $^{13}$CO(3-2) data will be discussed in this article.

\subsection{NANTEN2: $^{12}$CO(1-0) in Musca}
To obtain a more general view of the Musca cloud, we use $^{12}$CO(1-0) observations of the Musca cloud with the NANTEN2 telescope at Pampa la Bola in the Atacama desert. These observations were carried out with the single-beam SIS receiver on the telescope, developed by Nagoya University, with a 1 GHz bandwidth and a spectral resolution of 0.16 km s$^{-1}$ at 115 GHz. The observations made several 30$^{\prime}$ OTF maps, resulting in a full map size of 9 square degrees (7.3 pc x 7.3 pc at a distance of 140 pc) with an rms noise of 0.45 K within the spectral resolution of 0.16 km s$^{-1}$. This large map covers the Musca filament as well as the extended ambient cloud. The region of the Musca cloud covered by the NANTEN2 observations is indicated in Fig. \ref{columnDensityMap}. The main beam temperature scale was calibrated with Orion KL to be 52.6 K, such that the observations are consistent with the intensity of the CfA 1.2m telescope. The data was reduced with a linear baseline fit to the emission-free part of the spectrum.

\subsection{\textit{Herschel} and extinction maps}
The column density and dust temperature profile close to the filament are provided by data from the \textit{\textit{Herschel}} Gould Belt Survey (HGBS)\footnote{\url{http://gouldbelt-herschel.cea.fr/archives}} \citep{Andre2010,Cox2016}. For a more extended view on the mass distribution in the Musca cloud, we derived an extinction map of the region using the code {\it A$_{\rm v}$Map} \citep{Schneider2011} which measures the average reddening of background stars in the 2MASS catalogue. The map was smoothed to a resolution of 5$'$ in order to improve the quality of the extinction map. In Fig. \ref{columnDensityMap} the high resolution \textit{\textit{Herschel}} column density map (18.2$''$) is embedded in the extinction map. In order to combine the \textit{\textit{Herschel}} and large scale extinction data, a linear relation between the \textit{\textit{Herschel}} column density and extinction was fitted. A conversion N$_{H_{2}}$ = 0.83$\pm$0.02$\cdot$10$^{21}$ A$_{\rm v}$ gave the best fit, which allowed to convert extinction to column density based on the information in the \textit{Herschel} map. This conversion factor from extinction to column density for Musca is close to the canonical values reported for large samples in the galaxy \citep[e.g.][]{Bohlin1978,Guver2009,Rachford2009}.

\section{Results}

\begin{figure}
\includegraphics[width=\hsize]{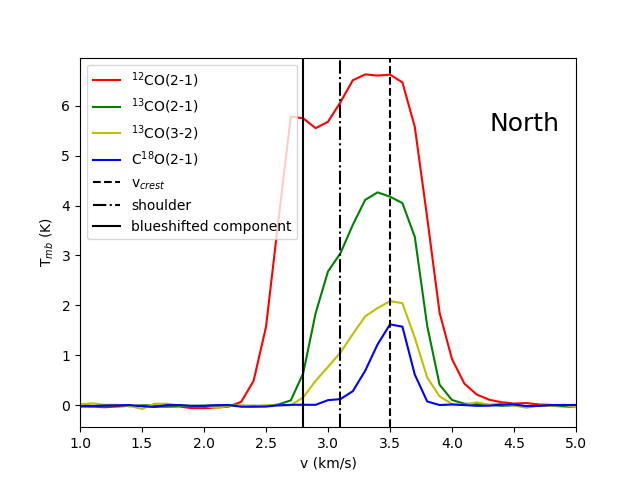}
\includegraphics[width=\hsize]{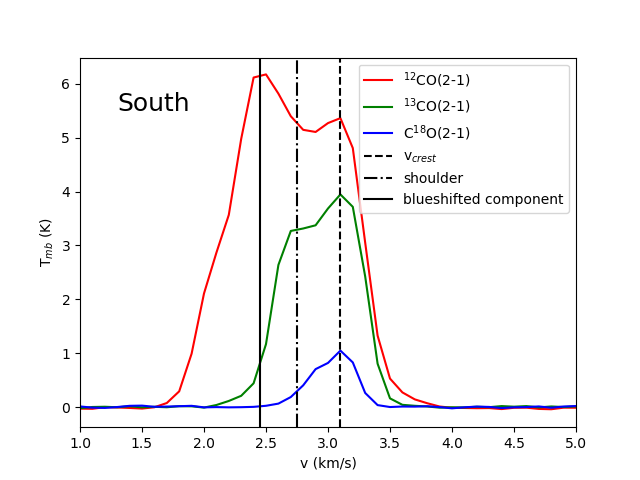}
\caption{\textbf{Top}: Averaged CO(2-1) isotopologue spectra and the $^{13}$CO(3-2) spectrum observed towards the filament crest (N$_{H_{2}} > $ 3$\cdot$21 cm$^{-2}$) in the northern map, showing the presence of the three components. \textbf{Bottom}: The same for the crest of the southern map.}
%, without $^{13}$CO(3-2) since it is not observed towards the southern map.}
\label{averageSpecNewComp}
\end{figure}

In this section, the first results from the CO(2-1) isotopologues are presented. The emission from these isotopologues in the northern and southern APEX map is presented in Figs. \ref{spectraMapNorth} and \ref{spectraMapSouth}, while the average spectra towards the filament crest are presented in Fig. \ref{averageSpecNewComp}. The observations will be presented step by step, starting with C$^{18}$O, followed by $^{13}$CO and in the end $^{12}$CO, because it can be observed in the before mentioned figures that the spectra show increasing complexity from C$^{18}$O to $^{12}$CO.

\subsection{C$^{18}$O emission from the filament crest}
\label{sec: c18osec}
\begin{figure}
\includegraphics[width=\hsize]{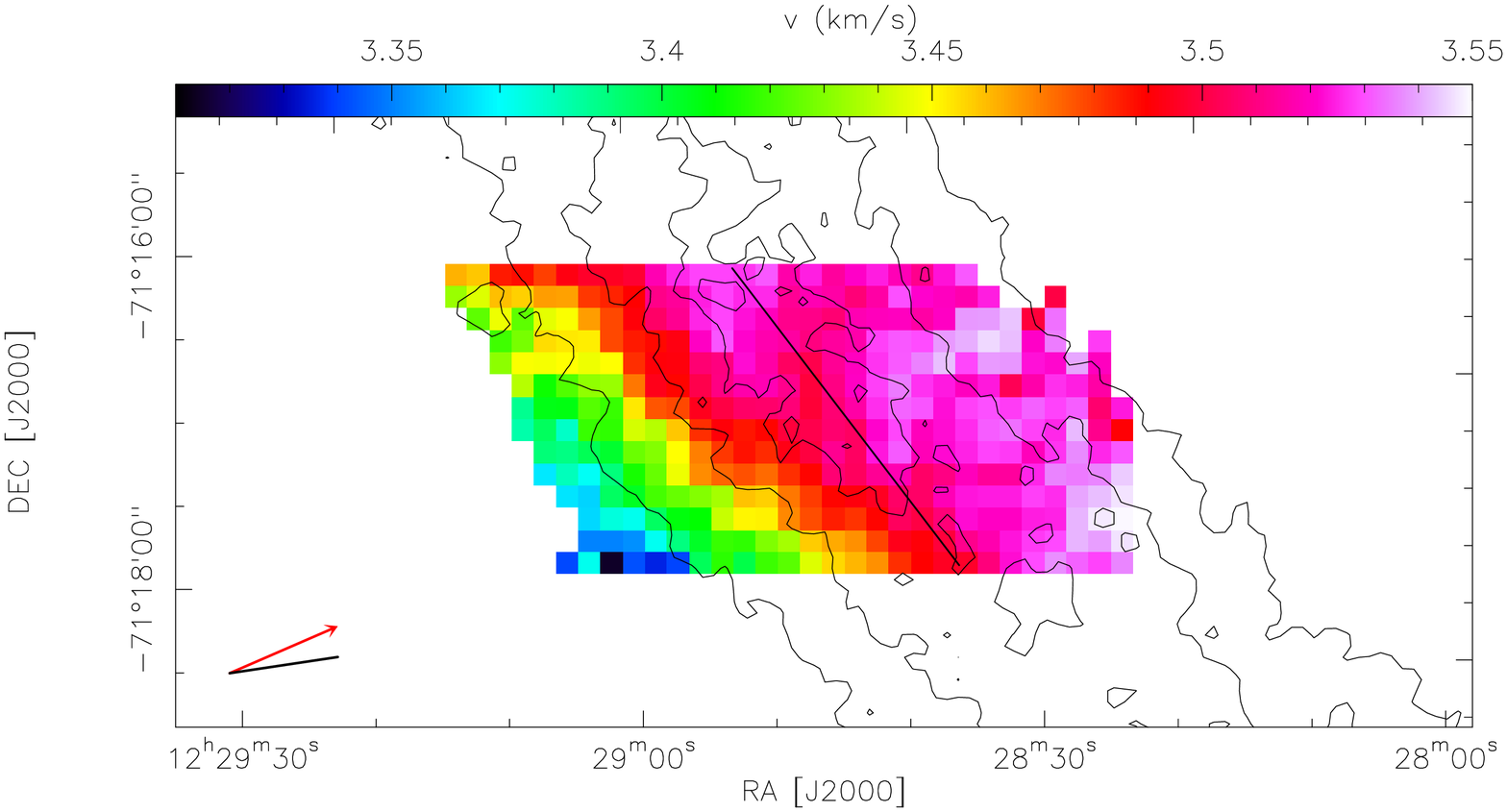}
\includegraphics[width=\hsize]{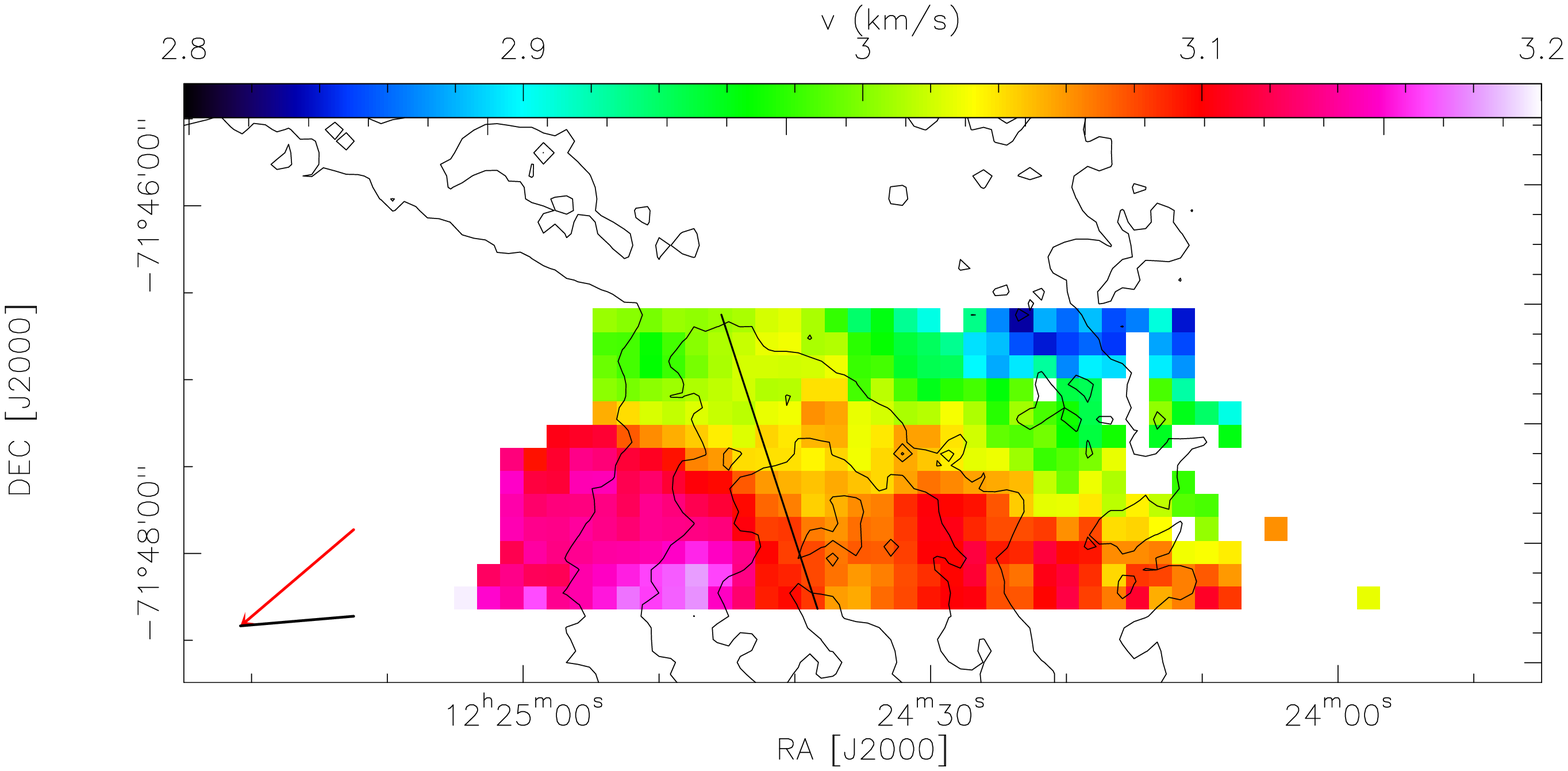}
\caption{Velocity field obtained over the Musca filament crest from the C$^{18}$O(2-1) line in the northern (top) and southern (bottom) map. The contours indicate the column density levels of the filament crest (N$_{H_{2}} =$ 3-6$\cdot$10$^{21}$ cm$^{-3}$). The black line on the velocity field connects the maximal column density at the top and bottom of the map, indicating the local orientation of the filament axis. In the lower left corner, the red arrow indicates the orientation of the velocity gradient, while the black line indicates the orientation of the magnetic field.}
\label{c18oVelFields}
\end{figure}

\begin{figure}
\includegraphics[width=\hsize]{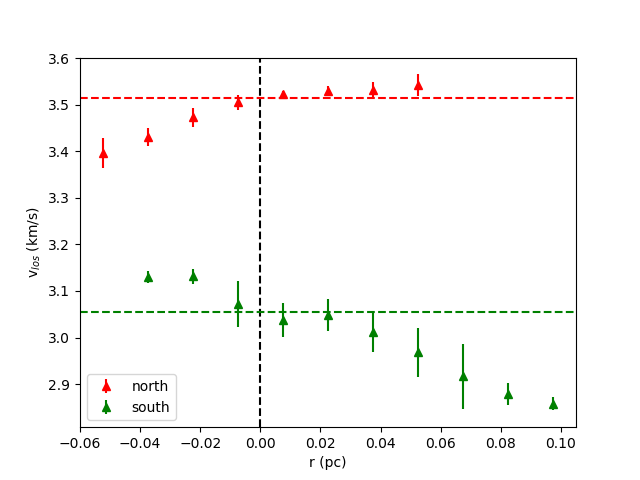}
\caption{Line-of sight velocity across the filament derived from C$^{18}$O(2-1) emission averaged along the filament as a function of the distance (r) for the northern (red) and southern (green) maps. The distance is always determined perpendicular to the local orientation of the filament crest center. The horizontal dashed lines indicate the velocity at the center of the filament crest, the vertical dashed line indicates the center of the crest, and the errorbars indicate the dispersion at each radius in the maps.}
\label{velVsRadius}
\end{figure}

Figures \ref{spectraMapNorth} and \ref{spectraMapSouth} show the $^{13}$CO(2-1) and C$^{18}$O(2-1) spectra of the northern and southern map overlaid on the \textit{Herschel} column density. In both maps, C$^{18}$O(2-1) is only detected at the crest of the Musca filament within a physical size of 0.1 - 0.15 pc, and the brightness of C$^{18}$O(2-1) quickly decreases when moving towards the strands around the filament crest. It suggests that C$^{18}$O is tracing only the highest column density regions of the filament (typically N$_{H_{2}}>$ 3$\times$10$^{21}$ cm$^{-3}$).\\ 
Inspecting the spectra in both maps, we confirm that the filament crest has a single velocity component with a transonic linewidth of 0.15 km s$^{-1}$, going up to 0.2 km s$^{-1}$, as reported in \cite{Hacar2016}. However, it should be noted that Fig. \ref{averageSpecNewComp} shows that the C$^{18}$O emission in the southern map has a pronounced shoulder which might be the result of a second velocity component. Additionally, we note a slight and continuous shift of the central velocity across the crest. To further investigate this velocity structure, a gaussian line profile was fitted to the C$^{18}$O(2-1) spectra above the 3$\sigma$ noise rms. The velocity field obtained from the fitting is presented in Figs \ref{c18oVelFields} and \ref{velVsRadius}. These figures demonstrate organised velocity fields across the crest in both maps with a typical velocity interval of $\sim$ 0.2 km s$^{-1}$, which is similar to the spectral linewidth, suggesting a part of the linewidth might be due to these gradients. The velocity gradients were calculated using the nearest neighbour values for every pixel. They have a magnitude of 1.6 and 2.4 km s$^{-1}$ pc$^{-1}$ and an angle of 77$^\circ$ and 45$^\circ$ compared to the local orientation of the filament crest for the northern and southern map, respectively. %The angle between the average velocity gradient over all pixels and the filament crest is  for the northern and southern map, respectively. 
One should note that the southern map covers a part of the filament that is fragmenting which might have an impact on the observed velocity gradient \citep[e.g][]{Hacar2011,Williams2018,Arzoumanian2018}. We also find that the central velocity at the crest is $\sim3.5\,$km s$^{-1}$ in the north and around 3.0-3.1 km s$^{-1}$ in the south.\\ %The C$^{18}$O(2-1) linewidth in both maps is around 0.15 km s$^{-1}$, going up to 0.2 km s$^{-1}$, as is reported in \cite{Hacar2016}.\\
Since the velocity gradient is in the opposite direction in the south compared to the north, see Fig. \ref{velVsRadius}, this does not lead to a straightforward analysis of the entire Musca filament crest as a simple rotating filament which is theoretically studied in \cite{Recchi2014}. %Similar behaviour of a changing velocity gradient orientation at different locations in a filamentary structure was also seen in the massive star forming DR21 ridge \citep{Schneider2010}, and is discussed in Sect. \ref{sec: massAccretion}.

\subsection{$^{13}$CO emission from the strands}
\label{sec: 13coResults}
\begin{figure}
\includegraphics[width=1.08\hsize]{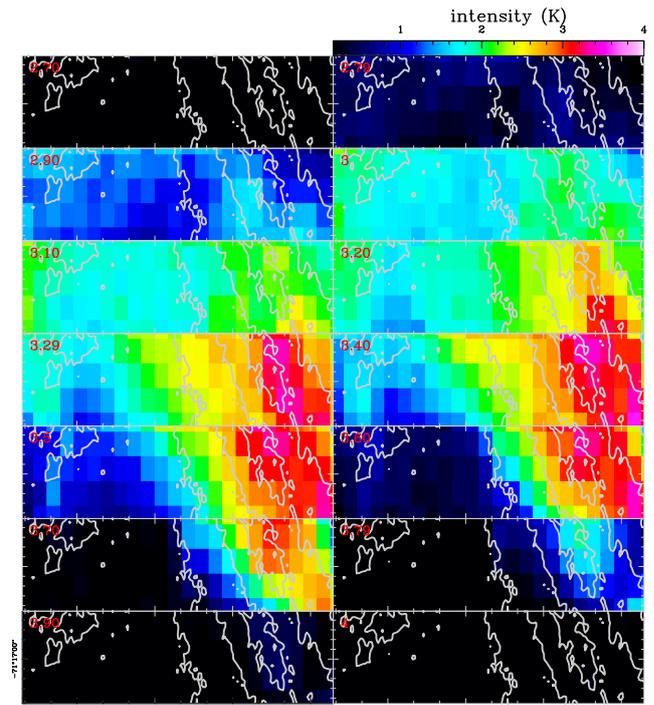}
\includegraphics[width=1.08\hsize]{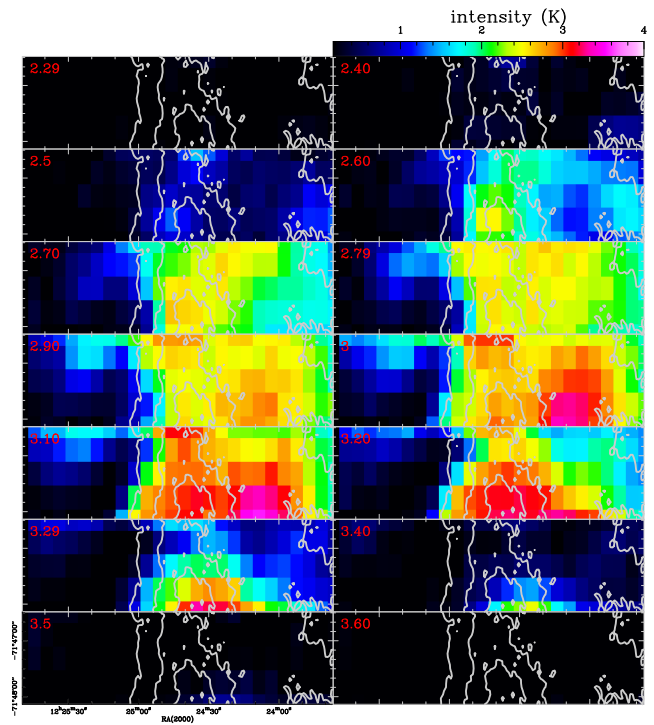}
\caption{\textbf{Top}: Channel maps of the APEX $^{13}$CO(2-1) observations in the northern map. The contours indicate the \textit{\textit{Herschel}} column densities N$_{H_{2}}$ = 2$\cdot$10$^{21}$, 3.5$\cdot$10$^{21}$ and 5$\cdot$10$^{21}$ cm$^{-2}$. The velocity of the channel is indicated in red at the upper left corner of every channel. It can be seen that the velocities of the blueshifted shoulder trace the strands. \textbf{Bottom}: The same for the southern map.}
\label{13coChannelmaps}
\end{figure}

In contrast to C$^{18}$O(2-1), the emission of $^{13}$CO(2-1) is not confined to the filament crest alone, but is also high towards the strands of Musca, see Figs. \ref{spectraMapNorth}, \ref{spectraMapSouth} and \ref{13coChannelmaps}. At the locations where the strands are not bright in the \textit{\textit{Herschel}} dust column density map, the $^{13}$CO emission also strongly decreases, see Figs. \ref{spectraMapSouth} and \ref{13coChannelmaps}.\\
Inspecting the $^{13}$CO(2-1) spectra of both maps in more detail, a component at the same velocity as the filament crest can be found, as well as a slightly blueshifted shoulder at v$_{los} \sim$ 3.1 km s$^{-1}$ in the northern map and v$_{los} \sim$ 2.7-2.8 km s$^{-1}$ in the southern map, see Fig. \ref{averageSpecNewComp}. The velocity component related to the filament crest disappears when moving away from the filament crest, while the emission from the shoulder remains present over the strands, see e.g. Figs. \ref{spectraMapNorth} and \ref{13coChannelmaps}. This implies that the observed shoulder comes from slightly more blueshifted emission related to the strands. From this moment on, when talking about the shoulder we thus implicitly talk about the strands and vice versa.\\\\
The velocity field of C$^{18}$O(2-1) demonstrated internal motion in the filament crest. The $^{13}$CO(2-1) spectra provide more information, namely that there are two velocity components: one related to the filament crest %\citep{Hacar2016} 
and a second component related to the strands.\\% A possible relation between the strand kinematics and the internal motion in the filament crest will be addressed in a later section.\\
At this point, we have presented the blueshifted shoulder that correlates well with the \textit{\textit{Herschel}} strands in both maps. This indicates that the strands around the filament crest are blueshifted compared to the filament crest. When inspecting the $^{13}$CO(2-1) channel maps of the southern map in Fig. \ref{13coChannelmaps} in more detail, one finds that east of the filament crest there is some weak $^{13}$CO(2-1) emission (T$_{mb} \lesssim$ 1 K) away from the filament crest at more or less the same velocity as the filament crest, see Figs. \ref{spectraMapSouth} and \ref{13coChannelmaps}. This emission suggests that at some locations there is also a small amount of $'$redshifted$'$ gas near the filament crest that is roughly at the same velocity along the line of sight as the filament crest itself.

\subsection{$^{12}$CO: Blueshifted emission}
Inspecting the APEX $^{12}$CO(2-1) spectra, another velocity component shows up with little corresponding emission in $^{13}$CO(2-1), see e.g. Figs. \ref{spectraMapSouth} and \ref{averageSpecNewComp}. This new velocity component, which is even more blueshifted than the shoulder, is observed both in the northern and southern map and also at the locations where the strands disappear (in the {\it Herschel} column density map), see Fig. \ref{spectraMapSouth}. This suggests, as $^{13}$CO(2-1) is only marginally detected at this velocity, that there is low column density gas present around the Musca filament with blueshifted velocities along the line of sight of 2.7 km s$^{-1}$ in the north and 2.5 km s$^{-1}$ in the south, see Fig. \ref{averageSpecNewComp}. 
We refer to it as $'$the blueshifted component$'$ from now on.\\
These observations of the CO isotopologues thus indicate that the Musca cloud has two more velocity components on top of the already established velocity-coherent filament crest: the shoulder (detected in $^{13}$CO) and the blueshifted velocity component (detected in $^{12}$CO). These different velocities can also be seen as a continuous velocity structure from large scale blueshifted gas to the small scale crest which corresponds to the reddest CO emission of the region. This generally indicates that the kinematics in the Musca cloud is more complex than the single velocity component of the filament crest \citep{Kainulainen2016,Hacar2016}, and that large scale kinematics might play an important role in the filament formation and cloud evolution.\\\\
%With observations of $^{12}$CO(2-1), $^{13}$CO(2-1) and C$^{18}$O(2-1) we have thus shown that: C$^{18}$O traces the crest, $^{13}$CO traces the strands and $^{12}$CO traces the more diffuse ambient cloud. 

\subsection{NANTEN2 $^{12}$CO(1-0) data}
\label{sec: NantenSection}
\begin{figure*}
\includegraphics[width=\hsize]{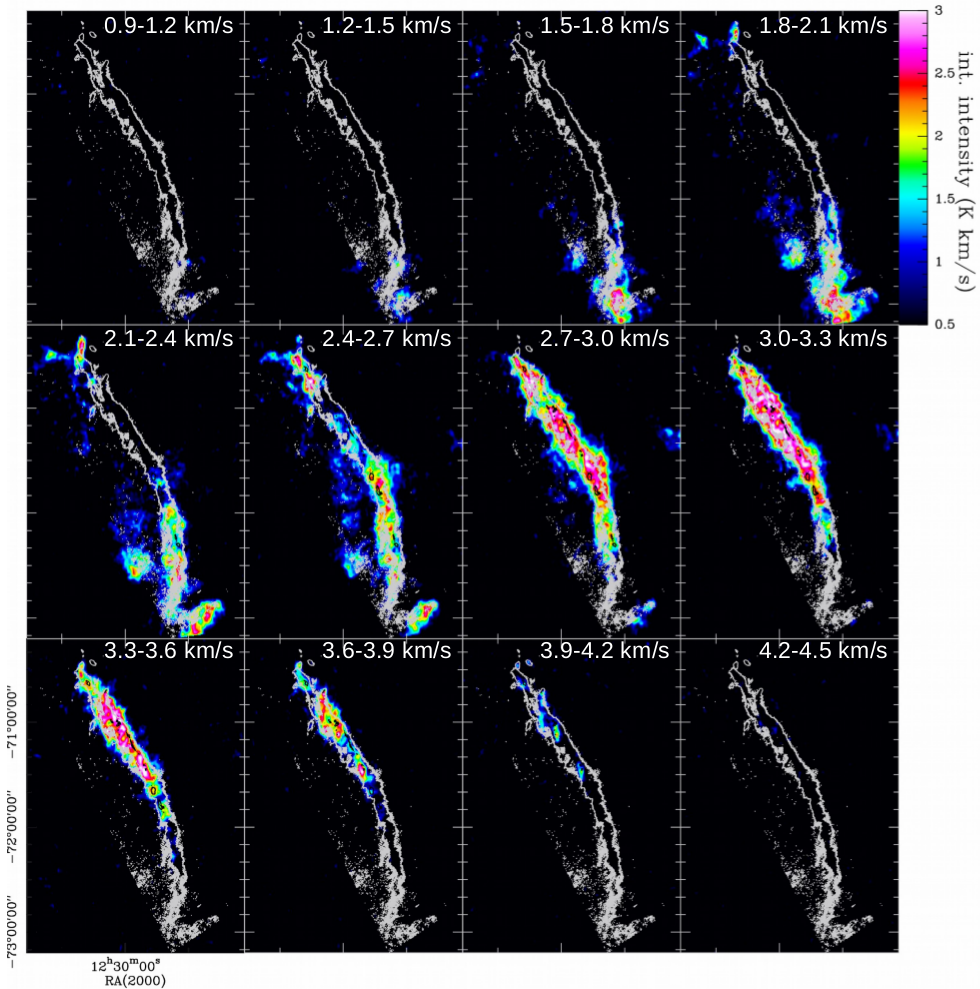}
\caption{Integrated intensity of NANTEN2 data over velocity intervals of 0.3 km s$^{-1}$. The black contours at N$_{H_{2}}$ = 5$\cdot$10$^{21}$ cm$^{-2}$ from the \textit{Herschel} data indicate the center of the filemant crest. These black contours are best visible at velocities between 2.7 and 3.6 km s$^{-1}$ when there is $^{12}$CO(1-0) emission towards the crest. The grey contour, at N$_{H_{2}}$ = 2$\cdot$10$^{21}$ cm$^{-2}$, indicates the area that encloses the strands. It can be observed that there is some extended emission east of the Musca filament and that the filament crest is not at the middle of the area containing the strands.
}
\label{largeMapNANTEN}
\end{figure*}
\begin{figure*}
\includegraphics[width=0.5\hsize]{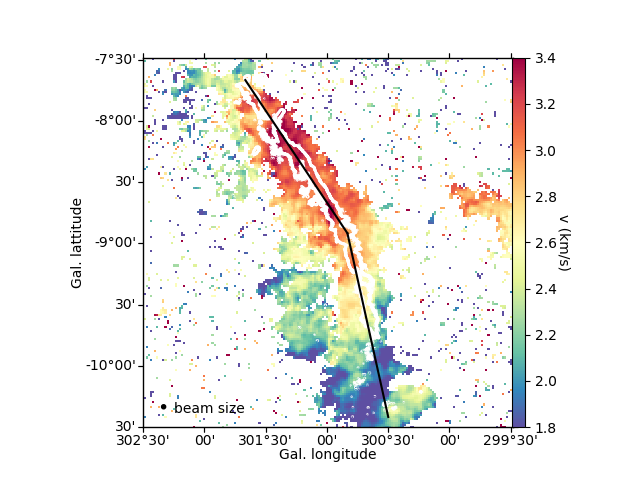}
\includegraphics[width=0.5\hsize]{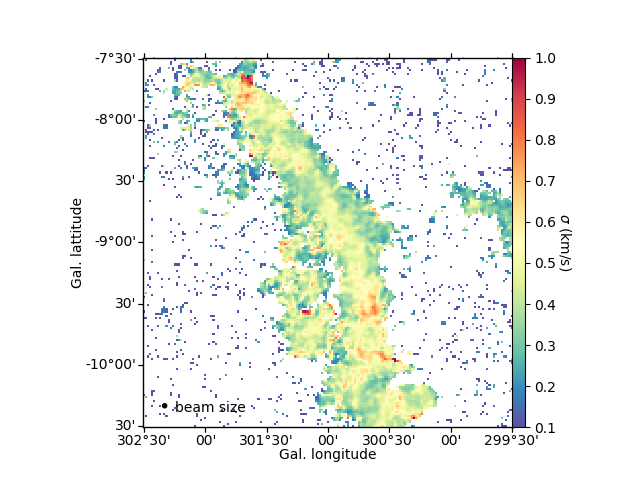}
\caption{\textbf{Left:} $^{12}$CO(1-0) velocity field in the Musca cloud from fitting a single gaussian to the $^{12}$CO(1-0) NANTEN2 data, showing an organised velocity field along the Musca filament. The black line indicates the central axis used to construct the PV diagram in Fig. \ref{PVdiagrams} perpendicular to the Musca filament. The white contours show the high column density region (N$_{H_{2}} >$ 2$\cdot$10$^{21}$) in Musca. \textbf{Right}: The $^{12}$CO(1-0) linewidth from the Gaussian fitting to the NANTEN2 data.}
\label{fitsNANTEN}
\end{figure*}
\begin{figure}
\includegraphics[width=\hsize]{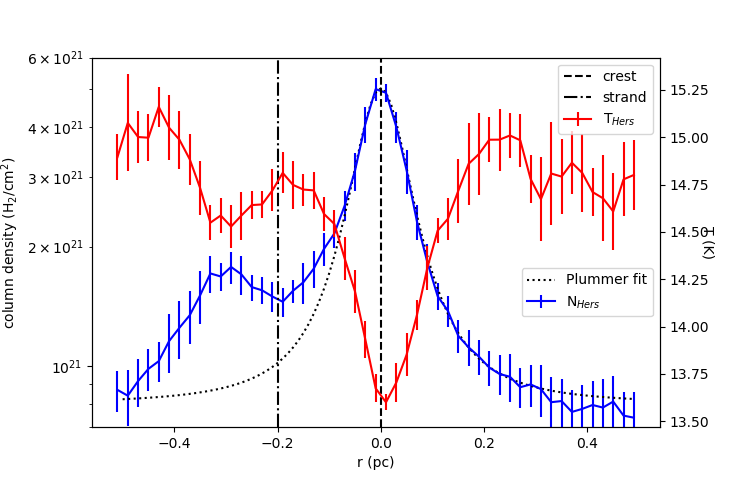}
\includegraphics[width=\hsize]{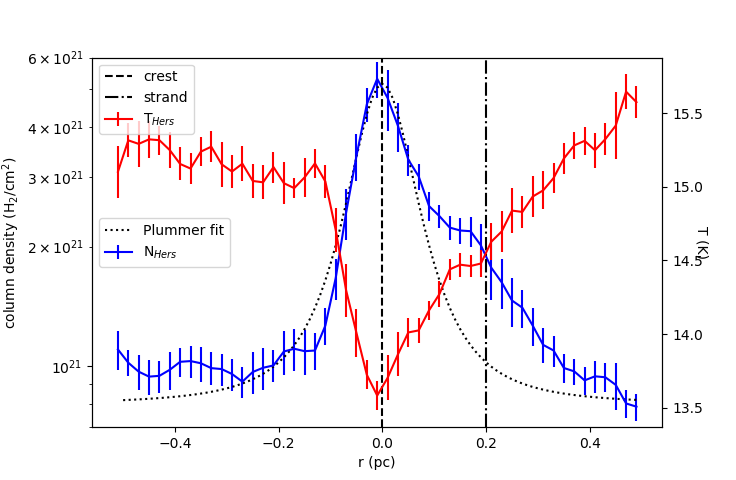}
\caption{\textbf{Top}: \textit{\textit{Herschel}} column density and temperature profile in the northern map. It demonstrates an asymmetry of the column density due to the strands. The fitted Plummer profile (excluding the strands) shows that the strand is a dense structure directly next to the filament crest. The standard deviation from the average value is also indicated at each radius. A negative radius was chosen to be located east of the filament crest. \textbf{Bottom}: The same for the southern map, note that the strand has changed side.}
\label{colDensProfile}
\end{figure}

The $^{12}$CO(1-0) mapping performed by the NANTEN2 telescope provides a view on the large scale kinematics of the Musca filament and the ambient cloud. Fig. \ref{largeMapNANTEN} presents channel maps of this data set, which shows that the brightest $^{12}$CO(1-0) gas is located at the crest and strands defined by the \textit{Herschel} data.\\
In the channel maps with v $<$ 2.7 km s$^{-1}$, there is also emission observed outside the filament contours. This is confirmed with the velocity field obtained from the NANTEN2 data in Fig. \ref{fitsNANTEN}, which demonstrates that the blueshifted component traces a more diffuse and extended ambient cloud that is observed in dust continuum emission with \textit{\textit{Herschel}} and in the extinction map.  At velocities between 2.7 and 4 km s$^{-1}$, corresponding to the velocity of the shoulder and crest, the emission is nicely constrained to the filament seen by \textit{\textit{Herschel}} as was already inferred from the APEX data, see Fig. \ref{largeMapNANTEN}.\\
The NANTEN2 observations confirm that the large scale, surrounding gas of the Musca filament is mostly blueshifted. It also confirms that smaller scale blueshifted strands can be found at both sides of the filament crest, as was already suggested from the APEX data. This indicates that there are local changes in the position of these blueshifted strands compared to the crest. The column density profile perpendicular to the filament crest in Fig. \ref{colDensProfile} also shows that the filament crest, both in the south and north, is not located at the center of the strands. It even shows that the crest has a direct border with the more diffuse ambient cloud at some locations, which results in locally asymmetric column density profiles.\\
Furthermore, NANTEN2 data shows that there is virtually no CO emission in the whole Musca region that is redshifted compared to the crest at 3.0 - 3.5 km s$^{-1}$. This confirms the tendency already noted at small scales with the APEX data in the two maps. %was already noted from the spectra obtained with APEX, but appears to be more generally the case at larger scales, see Fig. \ref{largeMapNANTEN}. 
%We indeed note that above a velocity of $\sim 3\,$km/s all the CO emission is confined inside the \textit{\textit{Herschel}} column density contours, i.e. inside the strands and crest parts of the filament. In contrast some significant blueshifted emission is found extended outside the filament, and mostly in the east, south-east, and south of the filament.\\
In conclusion, the Musca filament and its possible gas reservoir have an interesting asymmetric distribution both spatially (crest concentrated in the west/north-west direction) and kinematically (redshifted crest and strands compared to the blueshifted large-scale ambient cloud).
%In the ambient cloud, only a small amount of $'$redshifted$'$ CO gas (v $>$ 2.7 km s$^{-1}$) is found, which is located west of the filament (near $\delta_{\text{2000}}$ = -71$^{o}$20$^{\prime}$ in Fig. \ref{largeMapNANTEN}). It can be noted that the velocity of this gas is very similar to the velocity of the Musca filament crest (2.7-3.3 km s$^{-1}$). This suggests that the Musca filament crest might have an ambient cloud that is asymmetric in velocity space with a limited amount of redshifted ambient gas that is roughly at the same velocity as the filament crest along the line of sight. This also suggests that the large scale red- and blueshifted ambient cloud might be separated by the filament as in B211/3 \citep{Palmeirim2013}.

\subsection{Orientation of the velocity gradient and the magnetic field}
\label{sec: MagFieldVsGrad}
An organised velocity gradient over the Musca filament crest was reported in Sec. \ref{sec: c18osec} from C$^{18}$O observations. Here, the orientation of this velocity gradient is compared to the orientation of the local magnetic field. For the entire Musca filament, the magnetic field is nearly perpendicular to the filament crest \citep{Pereyra2004,PlanckArzoumanian2016}, with a typical angle between the filament crest and the magnetic field of 80$^{\circ}$ \citep{Cox2016}.\\\\
In order to further investigate a possible link between the plane of the sky magnetic field orientation and the velocity gradient in the filament crest, the magnetic field was constructed from the Planck 353 GHz data \citep{PlanckCollaboration2016data}. To reduce the noise in the magnetic field map, the I, Q and U maps were smoothed to a resolution of 10$'$ (or 0.4 pc for Musca at 140 pc) \citep{Planck2016Soler,PlanckArzoumanian2016}. The orientation of the magnetic field was then calculated in the IAU convention using the formulas presented in \cite{Planck2016Soler,PlanckArzoumanian2016}. In the area covered by the C$^{18}$O velocity maps, the magnetic field orientation \citep{PlanckArzoumanian2016} and the average velocity gradient were compared in Fig. \ref{c18oVelFields}, showing no clear correlation. In the northern map, the angle between the magnetic field orientation and the average velocity gradient is 28$^{\circ}$. In the southern map the angle is 54$^{o}$. A word of caution on this comparison: the velocity gradient is located over the filament crest, while the resolution of the Planck magnetic field orientation is significantly larger than the $\sim$ 0.1 pc size of the filament crest. It does however show that the velocity gradient over the filament crest has a significant offset compared to the large scale organised magnetic field in its close surroundings.\\ %This could indicate that the flows towards the Musca filament are not necessarily perfectly aligned with the magnetic field.\\

\section{Line radiative transfer analysis}
\label{sec: analysis}

\subsection{Density in Musca from radiative transfer of CO}
\label{sec: density}
\begin{figure}
\includegraphics[width=\hsize]{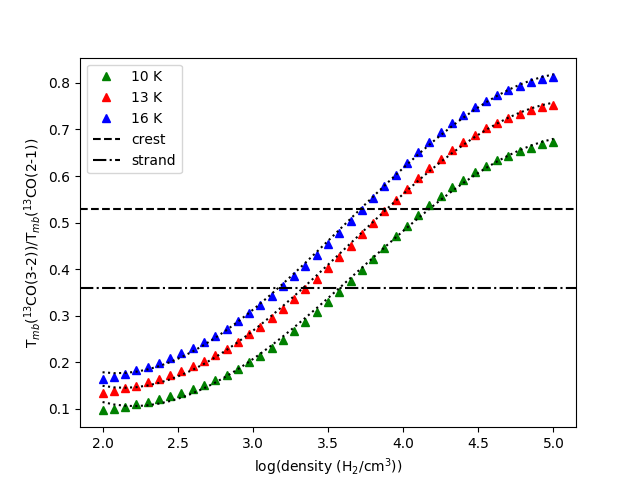}
\caption{Evolution of the $^{13}$CO(3-2)/$^{13}$CO(2-1) brightness temperature ratio as a function of density for realistic temperatures of the Musca filament. The dotted curves are the functions fitted through the RADEX results to invert the observed ratio in Musca to a density. The horizontal lines indicate the average observed ratios for the strand and the crest in the northern map.}
\label{ratT13covsDen}
\end{figure}

\begin{figure}
\includegraphics[width=\hsize]{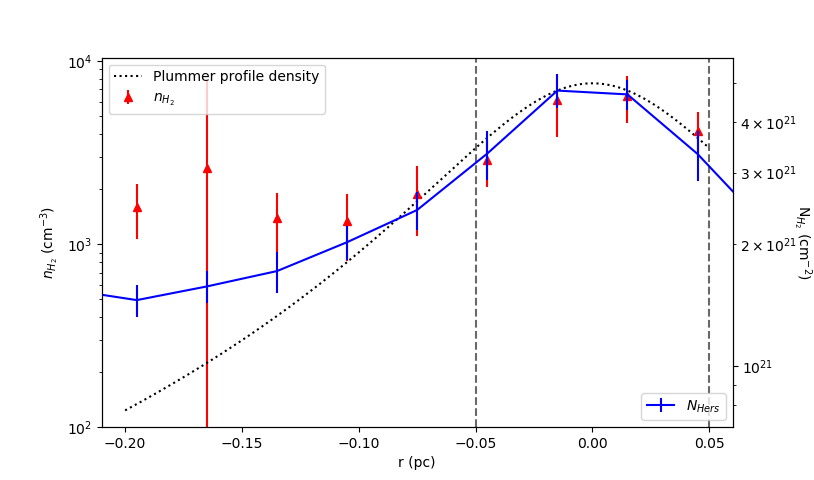}
\caption{In red, the density profile and its standard deviation obtained from the $^{13}$CO(3-2)/$^{13}$CO(2-1) ratio as a function of the distance from the filament crest for the northern map. The vertical dashed lines indicate the extent of the filament crest. It should be noted that the \textit{Herschel} temperature is too high for the filament crest such that the density will be slightly higher than the values shown in this figure, see Fig. \ref{ratT13covsDen}. The dotted black line describes the Plummer density profile for the Musca filament as derived from the column density profile using equation (1) in \cite{Arzoumanian2011}. This density profile fits with density predictions at the filament crest, but towards the strands the density is significantly higher than predicted from the Plummer profile. The \textit{Herschel} column density profile, in blue, is plotted on the axis on the right. When comparing the density profile with the column density profile, one should take into account the different scales of the left and right axes.}
\label{densFromHers}
\end{figure}

\begin{figure}
\includegraphics[width=\hsize]{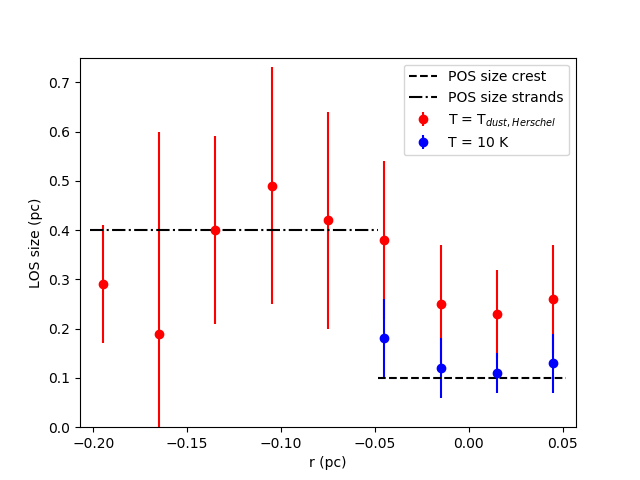}
\caption{Estimated characteristic size for the observed $^{13}$CO emission in the line of sight as a function of the distance (r) from the center of the filament crest. The characteristic size is estimated combining the \textit{Herschel} column density with the $^{13}$CO(3-2)/$^{13}$CO(2-1) density estimate using the \textit{Herschel} dust temperature (red) and a temperature of 10 K (blue). The horizontal lines indicate the characteristic sizes in the plane of the sky (POS) of the strands and the filament crest.}
\label{sizeLOS}
\end{figure}

With the non-LTE line radiative transfer code RADEX \citep{vanderTak2007}, which uses the LAMBDA database \citep{Schoier2005}, we investigate whether the CO isotopologue data can provide a consistent picture for the 3D geometry of the Musca filament. To do this, we begin with estimating the density profile across the Musca filament. In Fig. \ref{ratT13covsDen} it is shown that we can estimate the density with RADEX using the $^{13}$CO(3-2)/$^{13}$CO(2-1) brightness ratio, after smoothing $^{13}$CO(3-2) to the same resolution of $^{13}$CO(2-1) (here 28$^{\prime\prime}$), because this ratio strongly depends on the density. In particular between 10$^{2}$ - 5$\cdot$10$^{4}$ cm$^{-3}$, which covers the wide range of proposed densities at the Musca crest \citep{Kainulainen2016,Tritsis2018}, the ratio strongly depends on the density. 
We focus on the northern map since we only have $^{13}$CO(3-2) and $^{13}$CO(2-1) data for this location, which has a \textit{\textit{Herschel}} column density up to N$_{H_{2}} \sim$ 6$\cdot10^{21}$ cm$^{-2}$, see Fig. \ref{colDensProfile}.\\
In principle the line ratio does not depend on the abundance of $^{13}$CO, but since both $^{13}$CO lines become optically thick at the crest this is not completely true. However, in App. \ref{app: densityModelling} it is demonstrated that the varying optical depth related to the column density or abundance variations is not the main uncertainty for the $^{13}$CO(3-2)/$^{13}$CO(2-1) ratio.\\\\% does not strongly depend on the assumed column density and linewidth.\\\\% In \cite{Hacar2016} it was estimated that the opacity of $^{13}$CO(2-1) over the entire filament crest varies between 0.5 and 10, which fits with our calculations.\\
To construct the density profile, a RADEX grid of the $^{13}$CO brightness temperature ratio was created with 40 points on a log scale between n$_{H_{2}}$ = $10^{2}$ and $10^{5}$ cm$^{-3}$ for three different kinetic temperatures: 10, 13, and 16 K. This covers the range of \textit{Herschel} dust temperatures as well as kinetic gas temperatures put forward for Musca by \cite{Machaieie2017}. 
The FWHM and column density for $^{13}$CO that are used in the RADEX calculations are 0.7 km s$^{-1}$ and N$_{^{13}CO}$ = 1.1$\cdot10^{16}$ cm$^{-2}$, respectively, which is obtained from N$_{H_{2}}$ = 6$\cdot10^{21}$ cm$^{-2}$ using $\big[$H$_{2}\big]$/$\big[^{13}$CO$\big]$ = 5.7$\cdot10^{5}$ (from $\big[$H$_{2}\big]$/$\big[^{12}$CO$\big] \sim$ 10$^{4}$ and $\big[^{12}$CO$\big]$/$\big[^{13}$CO$\big] \sim$ 57 \citep{Langer1990}). The calculated opacities by RADEX for these models vary between 3 to 6 for $^{13}$CO(2-1) and 1 to 3 for $^{13}$CO(3-2), which fits with the estimated opacities in \cite{Hacar2016}.\\
To invert the observed ratios with APEX to a density profile, we work with the \textit{Herschel} temperature profile in Fig. \ref{colDensProfile}. %First, a function of the shape A + a$\cdot$x + b$\cdot$x$^{2}$ + c$\cdot$x$^{3}$ was fitted through the grid points of the RADEX calculations for each temperature to interpolate between the RADEX results, see Fig. \ref{ratT13covsDen}. The standard deviation of the fitted curve from the RADEX points varies between 0.0058 and 0.0071 for the three temperatures, which is good enough to confidently invert the observed ratio to a density, see Fig. \ref{ratT13covsDen}. Then, the density at different temperatures is interpolated on a log-scale to obtain the predicted density for the \textit{Herschel} dust temperature. 
The resulting density profile for the Musca filament is presented in Fig. \ref{densFromHers}.\\\\
This density profile predicts n$_{H_{2}}$ = 6.5$\pm$1.9$\cdot$10$^{3}$ cm$^{-3}$ at the filament crest that drops to n$_{H_{2}} \sim$ 1-3$\cdot$10$^{3}$ cm$^{-3}$ for the strands, see Fig. \ref{densFromHers}. However, for the interior of the filament crest the 13 K temperature is certainly overestimated. It was shown by \cite{Nielbock2012,Roy2014} that the temperature in the dense interior of different clouds is at least 3 K lower than the \textit{Herschel} dust temperature on the sky. Using a temperature of 10 K, which is actually also suggested by the LTE study of $^{13}$CO(1-0) and C$^{18}$O(1-0) by \cite{Machaieie2017} for the filament crest, one gets a more probable typical density in the crest of n$_{H_{2}}$ = 1.3$\pm$0.4 $\cdot$10$^{4}$ cm$^{-3}$.\\
Combining the obtained densities with the \textit{Herschel} column densities, one can estimate the size of the dense gas along the line of sight as a function of the distance from the filament crest. This is shown in Fig. \ref{sizeLOS}, demonstrating a typical size of 0.2-0.5 pc for the strands and a size of $\sim$ 0.25 pc at the filament crest for the overestimated temperature of 13 K. For the most probable 10 K temperature of the crest, we get a size of $\sim$ 0.1 pc for the crest. %However, taking into account a colder temperature of 10 K for the filament crest interiour, this points to a size along the line of sight of $\sim$ 0.1 pc. 
We thus get for both the filament crest and strands that their characteristic sizes along the lines of sight are roughly the same as their sizes in the plane of the sky, which implies a cylindrical geometry.\\\\
%Combining the density estimate with the \textit{\textit{Herschel}} column density of the crest (N$_{H_{2}} \sim$ 4$\cdot$ 10$^{21}$ cm$^{-2}$) and strands (N$_{H_{2}} \sim$ 10$^{21}$ cm$^{-2}$) it is possible to estimate a size along the line of sight for the crest and the strands. For the crest this gives a size of 0.1 pc and for the strands $\sim$ 0.1-0.3 pc. The crest thus has the same size along the line of sight as in the plane of the sky, and the strands have, within a factor 2-3, values similar to their size in the plane of the sky.\\\\
The density profile from the $^{13}$CO(3-2)/$^{13}$CO(2-1) brightness ratio, is compared in Fig. \ref{densFromHers} with the predicted density profile from fitting a Plummer profile to the filament crest in the northern APEX map \citep{Cox2016}. This shows that a Plummer profile provides an acceptable fit for the filament crest. In the strands we find that the density estimate from $^{13}$CO(3-2)/$^{13}$CO(2-1) is significantly higher than the density predicted by the Plummer profile, similar to what is observed for the column density in Fig. \ref{colDensProfile}. This suggests that the strands consist of denser gas than the surroundings, and it reinforces our earlier proposal that the strands are additional components on the top of the density (presumably plummer-like) profile of the filament crest.\\\\
We also verified the overall consistency of the obtained density results for the filament crest and strands by studying the observed brightness of the CO isotopologues, see Tab. \ref{brightnessRADEX}. For the filament crest, we find that n$_{H_{2}}$ = 10$^{4}$ cm$^{-3}$ and T$_{k}$ = 10 K reproduce the observed CO isotopologue brightnesses, even when taking into account a 30 \% uncertainty of the CO abundance as is shown in Tab. \ref{brightnessRADEX}. For the strands, we reproduce the observed $^{13}$CO brightness with n$_{H_{2}}$ = 10$^{3}$ - 3$\cdot$10$^{3}$ cm$^{-3}$ and T$_{k}$ = 15 - 18 K, however it is not possible to reproduce the C$^{18}$O brightness of the strands with typical [$^{13}$CO]/[C$^{18}$O] ratios around 10. This will be addressed in more detail in the next section.\\
These results suggest that the kinetic temperature of the strands and the filament crest are similar to the \textit{Herschel} dust temperature. Generally speaking it is found that T$_{dust} \le$ T$_{k}$ \citep[e.g.][]{Goldsmith2001,Merello2019}, but in the dense interior of the Musca cloud T$_{dust} \approx$ T$_{k}$ can be expected \citep[e.g.][]{Goldsmith2001,Galli2002,Ivlev2019}. In the simulations by \cite{Seifried2016}, the dust is not yet fully coupled to the gas at the predicted densities inside the filament crest (n$_{H_{2}} \sim$ 10$^{4}$ cm$^{-3}$), but in these simulations the dust temperature reaches values below 10 K.% such that a gas temperature of 10 K for Musca filament crest remains consistent.

%\begin{table*}
%\tiny
%\begin{tabular}{c|c|ccccccccc}
%r (pc) & T & -0.195 & -0.165 & -0.135 & -0.105 & -0.075 & -0.045 & -0.015 & 0.015 & 0.045\\
%\hline
%size LOS (pc) & T$_{d, Hers}$ & 0.29 $\pm$ 0.12 & 0.19 $\pm$ 0.41 & 0.40 $\pm$ 0.19 & 0.49 $\pm$ 0.24 & 0.42 $\pm$ 0.22 & 0.38 $\pm$ 0.16 & 0.25 $\pm$ 0.12 & 0.23 $\pm$ 0.09 & 0.26 $\pm$ 0.11\\
%size LOS (pc) & 10 K &  &  &  &   & & 0.18 $\pm$ 0.08 & 0.12 $\pm$ 0.06 & 0.11 $\pm$ 0.04 & 0.13 $\pm$ 0.06\\
%\end{tabular}
%\caption{The estimated characteristic size in the line of sight for the gas that is traced by $^{13}$CO as a function of the distance (r) from the center of the filament crest in the northern map. This size is estimated using the \textit{Herschel} column density and the $^{13}$CO(3-2)/$^{13}$CO(2-1) density profile at the \textit{Herschel} dust temperature (first row) and a temperature of 10 K (second row). The characteristic sizes are larger for the area associated with the strand (r $<$ -0.05 pc) than for the area associated with the filament crest (r $>$ -0.05 pc).}
%\label{losSizesTable}
%\end{table*}

\subsection{C$^{18}$O abundance drop in the strands}
\begin{figure}
\includegraphics[width=\hsize]{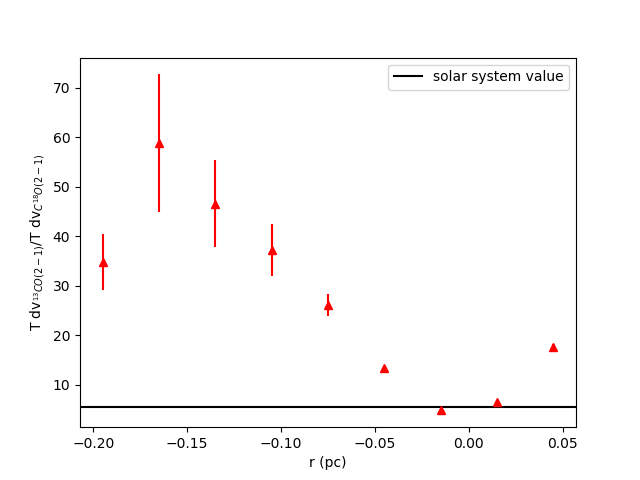}
\caption{Evolution of the $^{13}$CO(2-1)/C$^{18}$O(2-1) integrated brightness ratio as a function of distance from the filament crest in the northern map, showing a rapid increase as a function of the distance. It shows the average ratio for the values in each distance bin of 0.03 pc with the uncertainty related to the noise. %The grey area demonstrates the reported [$^{13}$CO]/[C$^{18}$O] ratio interval in Orion-A \citep{Shimajiri2014}.
}
\label{c18oVs13co}
\end{figure}

Studying the CO isotopologue emission in the strands, a strong variation of the [$^{13}$CO]/[C$^{18}$O] abundance ratio is observed. As already noted, the C$^{18}$O(2-1) emission, with a typical rms of $\sim$ 0.07 K, rapidly decreases in the strands until it is no longer clearly detected for a single beam in the strands. When studying the data in bins of 0.03 pc as a function of the distance from the center of the filament crest, we always obtain a $>$3$\sigma$ detection of C$^{18}$O(2-1). This then allows to follow the rapid continuous increase of [$^{13}$CO]/[C$^{18}$O] in Fig. \ref{c18oVs13co} as a function of the distance from the filament crest, showing that the ratio varies by an order of magnitude over less than 0.2 pc. Fig. \ref{c18oVs13co} displays the evolution of the integrated line brightness ratio as a function of the distance, which is a decent proxy for the $\big[^{13}$CO$\big]$/$\big[$C$^{18}$O$\big]$ ratio because of the very similar excitation conditions for both lines. This strong increase of the [$^{13}$CO]/[C$^{18}$O] ratio agrees with a weak C$^{18}$O(2-1) detection at the velocity of the strands with a brightness of $\sim$ 0.08 K when smoothing over the area covered by the strands, while the brightness of $^{13}$CO(2-1) in the strands is still $\sim$ 2.5 K, see Tab. \ref{brightnessRADEX}.\\ 
It was shown by \cite{Zielinsky2000} that a FUV field can create a variation of $\big[^{13}$CO$\big]$/$\big[$C$^{18}$O$\big]$ as a function of the column density due to fractionation \citep[e.g.][]{vanDishoeck1988}. With theoretical studies of gas exposed to a weak FUV-field ($\sim$ 1 G$_{\text{0}}$), it was shown that carbon fractionation reactions can increase the $^{13}$CO abundance up to a factor 2-3 and thus affect the CO isotopologue ratios, specifically at low A$_{\text{V}} \sim$ 1 \citep[e.g.][]{Visser2009,Roellig2013,Szucs2014}. For C$^{18}$O, the wavelengths of the dissociation lines can penetrate deeper into the interior of the molecular cloud because of its poor self-shielding \citep[e.g.][]{Glassgold1985}. Furthermore, oxygen fractionation reactions can significantly decrease the C$^{18}$O abundance which can further increase the [$^{13}$CO]/[C$^{18}$O] ratio. However, to this point, oxygen fractionation reactions have received little attention \citep{Loisin2019}.\\
The observations of Musca indicate that the weak FUV field in Musca ($<$ 1 G$_{0}$: see Paper II) might be able to significantly suppress the C$^{18}$O abundance in the strands to $\big[^{13}$CO$\big]$/$\big[$C$^{18}$O$\big]$ = 40-50 on average, see Tab. \ref{brightnessRADEX}. Inside the filament crest we find $\big[^{13}$CO$\big]$/$\big[$C$^{18}$O$\big] \lesssim$ 10. Significantly increasing $\big[^{13}$CO$\big]$/$\big[$C$^{18}$O$\big]$ at the crest would drop the brightness of C$^{18}$O(2-1) below the observed value. This confirms that the $\big[^{13}$CO$\big]$/$\big[$C$^{18}$O$\big]$ abundance ratio increases by roughly an order of magnitude over a small physical distance ($\le$ 0.2 pc in the plane of the sky) when going from the crest to the strands at lower A$_{\text{V}}$. In the study by \cite{Hacar2016} of the filament crest it was already noted that there were some indications of an increase of $\big[^{13}$CO$\big]$/$\big[$C$^{18}$O$\big]$ towards lower A$_{\text{V}}$ ($<$ 3) areas of the filament crest. Including the strands into the analysis, we now demonstrate that there is indeed a strong drop of C$^{18}$O abundance for low A$_{\text{V}}$ gas in the Musca cloud. This observation indicates that even a weak FUV field can have a large impact on the molecular cloud chemistry of CO. %as the observed $\big[^{13}$CO$\big]$/$\big[$C$^{18}$O$\big]$ variation for Musca is larger than found e.g. in the Orion-A molecular cloud \citep{Shimajiri2014,Ishii2019}, see Fig. \ref{c18oVs13co}.
The formation of the dense filament crest is thus necessary for sufficient shielding such that C$^{18}$O can form and implies that C$^{18}$O is not a good column density tracer.
\begin{table*}
\small
\begin{tabular}{l|ccccc|ccc}
% & observed & RADEX & observed & RADEX & observed & RADEX & observed & RADEX\\
 crest & N$_{\text{H}_{2}}$ (cm$^{-2}$) & T$_{kin}$ (K) & n$_{\text{H}_{2}}$ (cm$^{-3}$) &  [H$_{2}$]/[$^{13}$CO] & [$^{13}$CO]/[C$^{18}$O] & T$_{^{13}CO(2-1)}$ (K) & T$_{^{13}CO(3-2)}$ (K) & T$_{C^{18}O(2-1)}$ (K)\\
 \hline
  observed values &  &  &  & & & 4.2 & 2.1 & 1.6\\
 \hline
% crest & 4e21 & 9 & 1e4 & 0.6 & 3.7 & 1.5 & 1.3\\
% crest & 4e21 & 9 & 2e4 & 0.6 & 4.0 & 1.9 & 1.5\\
  & 4$\cdot$10$^{21}$ & 10 & 10$^{4}$ &  5.7$\cdot$10$^{5}$ & 7.3 & 4.6 & 2.1 & 1.6\\
  & 6$\cdot$10$^{21}$ & 10 & 10$^{4}$ &  5.7$\cdot$10$^{5}$ &  7.3 & 4.9 & 2.6 & 2.3\\
  & 6$\cdot$10$^{21}$ & 9 & 10$^{4}$ &  5.7$\cdot$10$^{5}$ & 7.3 & 4.1 & 2.0 & 2.0\\
 RADEX & 4$\cdot$10$^{21}$ & 11 & 10$^{4}$ &  8$\cdot$10$^{5}$ & 7.3 & 4.8 & 2.2 & 1.5\\
 %f3 & 4e21 ($\big[^{13}$CO$\big]$/$\big[$C$^{18}$O$\big]$=15) & 9 & 1e4 & 0.6 &  &  & 0.71\\
 %f4 & 4e21 & 9 & 1e4 & 0.4 &  &  & 1.8\\
 %f5 & 4e21 ($\big[^{13}$CO$\big]$/$\big[$C$^{18}$O$\big]$=15) & 9 & 1e4 & 0.4 &  &  & 1.0\\
 & 4$\cdot$10$^{21}$ & 13 & 6.5$\cdot$10$^{3}$ &  5.7$\cdot$10$^{5}$ & 7.3 & 6.1 & 3.0 & 1.9\\
 & 3$\cdot$10$^{21}$ & 10 &  10$^{4}$ &  5.7$\cdot$10$^{5}$ & 7.3 & 4.3 & 1.8 & 1.3\\
 %crest: 20 K & 4$\cdot$10$^{21}$ & 20 & 3$\cdot$10$^{3}$ & 0.6 & 6.9 & 3.3 & 1.8\\
 %strand & 1e21 & 15 & 4e3 & 0.6 & 2.7 & 0.87 & 0.48\\
 \end{tabular}
 \newline
 \vspace*{0.5 cm}
 \newline
 \begin{tabular}{l|ccccc|ccc}
 strand & N$_{\text{H}_{2}}$ (cm$^{-2}$) & T$_{kin}$ (K) & n$_{\text{H}_{2}}$ (cm$^{-3}$) &  [H$_{2}$]/[$^{13}$CO] & [$^{13}$CO]/[C$^{18}$O] & T$_{^{13}CO(2-1)}$ (K) & T$_{^{13}CO(3-2)}$ (K) & T$_{C^{18}O(2-1)}$ (K)\\
 \hline
 observed values &  &  &  & & & 2.4 & 0.8 & 0.085\\
 \hline
  & 10$^{21}$ & 15 & 3$\cdot$10$^{3}$  & 5.7$\cdot$10$^{5}$ & 7.3 & 2.8 & 0.85 & 0.52\\
  & 10$^{21}$ & 15 & 2$\cdot$10$^{3}$  & 5.7$\cdot$10$^{5}$ & 7.3 & 2.3 & 0.59 & 0.42\\
  & 10$^{21}$ & 18 & 2$\cdot$10$^{3}$  & 5.7$\cdot$10$^{5}$ & 7.3 & 2.6 & 0.8 & 0.48\\
 RADEX & 10$^{21}$ & 15 & 3$\cdot$10$^{3}$  & 3.8$\cdot$10$^{5}$ & 7.3 & 3.7 & 1.2 & 0.78\\
  & 10$^{21}$ & 15 & 10$^{3}$  & 3.8$\cdot$10$^{5}$ & 7.3 & 2.0 & 0.44 & 0.39\\
  & 10$^{21}$ & 15 & 10$^{3}$  & 2.9$\cdot$10$^{5}$ & 7.3 & 2.4 & 0.59 & 0.52\\
  & 10$^{21}$ & 15 & 10$^{3}$  & 1.9$\cdot$10$^{5}$ & 7.3 & 3.0 & 0.84 & 0.74\\
 & 10$^{21}$ & 15 & 3$\cdot$10$^{3}$ &  5.7$\cdot$10$^{5}$ & 40 & 2.8 & 0.85 & 0.10\\
 & 10$^{21}$ & 15 & 3$\cdot$10$^{3}$ &  5.7$\cdot$10$^{5}$ & 50 & 2.8 & 0.85 & 0.08
\end{tabular}
\caption{{\bf Top table:} Results of several RADEX calculations for the filament crest using a FWHM of 0.5 km s$^{-1}$, and with varying column densities, temperatures and [H$_{2}$]/[$^{13}$CO] abundance ratios. The predictions closely match the observed brightness towards Musca filament crest for n$_{H_{2}}$ = 10$^{4}$ cm$^{-3}$ and T$_{k}$ = 10 K, in particular when working with N$_{H_{2}}$ = 4$\cdot$10$^{21}$ cm$^{-2}$ which is the column density that is associated with filament crest. The presented observed brightness is obtained from the average spectra at N$_{H_{2}} >$ 3$\cdot$10$^{21}$ cm$^{-2}$. {\bf Bottom table:} RADEX calculations for the strands  with varying abundances for $^{13}$CO and C$^{18}$O. The [H$_{2}$]/[$^{13}$CO] abundance ratio is varied up to a factor 3, supported by theoretical models, and the [$^{13}$CO]/[C$^{18}$O] abundance ratio is varied to obtain results consistent with the observations. It demonstrates a high average [$^{13}$CO]/[C$^{18}$O] abundance ratio ($\sim$ 50) in the strands. The presented observed brightness is obtained from the average spectra at N$_{H_{2}} <$ 3$\cdot$10$^{21}$ cm$^{-2}$. The observed brightness towards the strands generally matches predictions using n$_{H_{2}}$ = 10$^{3}$-3$\cdot$10$^{3}$ and T$_{K}$ = 15-18 K. %The results for a model with n$_{H_{2}}$ = 6.5$\cdot$10$^{3}$ cm$^{-3}$ and T$_{k}$ =  13 K is also presented, showing that it does not match with the observed brightness of the filament crest. Bottom table: % We took into account a column density of N$_{H_{2}} \sim$ 10$^{21}$ cm$^{-2}$ associated with the ambient cloud, we use N$_{H_{2}}$ = 10$^{21}$ cm$^{-2}$ for the strands and N$_{H_{2}}$ = 4$\cdot$10$^{21}$ cm$^{-2}$ for the filament crest. $\big[$H$_{2}\big]$/$\big[^{13}$CO$\big]$ is taken to be 5.7$\cdot$10$^{5}$, unless specified otherwise in the first column. The calculation with $\big[$H$_{2}\big]$/$\big[^{13}$CO$\big]$ = 8$\cdot$10$^{5}$ was done to check the influence of possible abundance uncertainties (in this case 30\%), showing that an abundance variation has a small impact on the gas temperature. %The obtained values from RADEX for $^{13}$CO(2-1), $^{13}$CO(3-2) and C$^{18}$O(2-1) are the main beam brightness. The observed main beam brightness in the north for $^{13}$CO(2-1), $^{13}$CO(3-2) and C$^{18}$O(2-1) at the filament crest is: 4.2, 2.1 and 1.6 K, respectively. At the strands the main beam brightness is 2.4, 0.8 and 0.085 K, respectively.
}
\label{brightnessRADEX}
\end{table*}

\section{Discussion}

%{\bf In the previous sections, the CO isotopologue kinematics of the Musca filament and ambient cloud were presented. These CO observations were also used to construct a density profile for the Musca filament and strands. In this section, the presented results will be combined to discuss the physical conditions and kinematic evolution of the Musca cloud. In particular, we aim to understand the constraints by these observations on the physics of the Musca filament formation.}

\subsection{A cylindrical geometry for the Musca filament crest and strands}
\label{sec: physCondSection}

In Sect.~\ref{sec: analysis}, from $^{\rm 13}$CO line ratios, we derived estimates of the densities for the emitting CO gas in a section of the Musca filament to be n$_{H_{2}}$ = $1-3$ and $6-13 \cdot 10^{3}\,$cm$^{-3}$ in the strands and crest respectively. Using the \textit{\textit{Herschel}} total column densities towards the same directions (crest and strands), and assuming the standard CO abundance and gas/dust ratio, we then obtained the typical sizes on the line of sight for the crest and strands: $0.1-0.2\,$pc for the crest and $0.2-0.5\,$pc for the strands. This is very similar to their respective sizes projected on the sky. This clearly shows that both the strands and the crest are not more elongated along the line of sight than their transverse size in the plane of sky. Our CO observations therefore confirm that the Musca crest and strands correspond more to a filament, that is a cylindrical structure, than to a sheet seen edge-on as proposed in \citet{Tritsis2018}. The $0.1-0.2\,$pc line of sight size of the crest is clearly in disagreement with the proposed 6 pc size in \citet{Tritsis2018} to explain the regular pattern of striations observed in the Musca cloud with MHD waves. 

Looking carefully at the \textit{\textit{Herschel}} maps in \cite{Cox2016} we actually see that the striations are not necessary originating from the crest of the filament as assumed in \citet{Tritsis2018}. The striations are indeed seen over an extended region of $\sim 2 \times 4\,$pc around the crest (see Fig.~2 in \citet{Cox2016}), that we here refer to as the ambient cloud. It could thus come from some extended, more diffuse gas around the filament. %In the 3D model used by \citet{Tritsis2018}, the crest and the more extended emission are supposed to have the same 3D shape meaning that if the striations seen in the extended emission is proposed to be elongated along the line of sight over 6 pc, the model imposes by construction that the crest has the same elongation. Here we believe that the crest can have a different 3D shape than the more diffuse cloud in which it is embedded. 

For the immediate surroundings of the crest seen in our APEX maps, we also get a density from CO which points to sizes along the line of sight not larger than $\sim 0.5\,$pc only. This is still far from the 6 pc required to explain the striations with the magnetohydrodynamic waves. On the other hand our APEX CO study is limited to only the most nearby regions of the crest, of the order of $\sim$0.2 pc. At larger scale, the ambient cloud surrounding the filament could have a line of sight size up to 6 pc, and could then host the striation pattern. Using the typical observed column density of N$_{H_{2}}$ = $1-3 \cdot 10^{21}\,$cm$^{-2}$ in the direction of the strands, and assuming all this column density could be due to the surrounding ambient cloud, we would get for a line of sight length of 6 pc a typical density for this medium as low as $\sim 50-150\,$cm$^{-3}$. Such low densities might actually not be visible in CO and may correspond to some surrounding CO dark gas.

In conclusion, our CO observations clearly show that the densest region (crest and strands) of Musca is mostly a cylindrical filamentary structure. The striations are either not related to the proposed MHD vibrational modes of \citet{Tritsis2018} or these modes, if they require a depth of 6 pc, have developed in the extended ambient, probably CO-dark low density medium ($\sim 50-150\,$cm$^{-3}$) surrounding the Musca filament.

\subsection{Continuous mass accretion on the filament crest in Musca}
\label{sec: massAccretion}
%\subsubsection{{\bf The velocity gradient and mass accretion}}
The velocity structure of the Musca filament crest, displayed in Fig. \ref{c18oVelFields} and that of the surrounding strands, traced with $^{13}$CO(2-1), clearly indicates that in both APEX maps velocity gradients from blue-shifted extended gas to red-shifted dense gas are monotonic. These continuous gradients are a hint of slowing down of the large scale gas reaching the crest, and then point to accretion. %For the densest regions at the crest, the observed velocity gradients in C$^{18}$O(2-1) could then point to angular momentum deposition onto the dense part of the filament.

%{\bf demonstrates that in both APEX maps velocity gradients are monotonic over the crest. These continuous gradients can be a hint of slowing down from the large scale gas reaching the crest, and can thus point to accretion. %such as for B211 \citet{Shimajiri2018}. 
%The observed velocity gradients in C$^{18}$O(2-1) could also be interpreted as rotational motion.
%Because the Musca filament crest is velocity-coherent without multiple fibers \citep{Hacar2016}, %finding an interpretation for the observed velocity gradients over the filament crest should be more straightforward than in other filaments \citep[e.g.][]{Dhabal2018}. 
%the velocity gradients might be a hint of some rotation like motion of the filament crest. }

\begin{figure}
\includegraphics[width=\hsize]{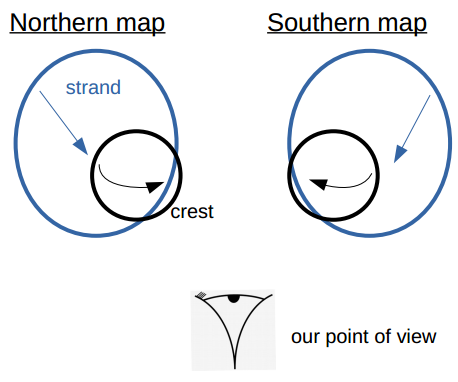}
\caption{Schematic illustration of the proposed accretion scenario responsible for the velocity gradient over the crest, possibly also related to some angular momentum deposition due to an accretion impact parameter, as seen from above the filament. The blue arrow indicates the motion of the strands compared to the crest from our point of view and the black arrows indicate the possible velocity field over the crest which is responsible for the observed velocity gradient.}
\label{sketchAccretion}
\end{figure}

Interestingly enough these monotonic gradients have an opposite direction at both observed locations, see Fig. \ref{c18oVelFields}, similar to what was observed for instance in the massive DR21 ridge \citep{Schneider2010}. In DR21 this behaviour was proposed to be the result of a global collapse with inflowing subfilaments driving the velocity gradients.\\
Studying the location and velocity of the strands in the observed APEX maps, we found that the strands have changed side of the filament from our point of view, see Fig. \ref{colDensProfile}, while at both locations the velocity of the strands are blueshifted, see e.g. Fig. \ref{13coChannelmaps}. 
The blueshifted part of the filament crest velocity field, traced by C$^{18}$O(2-1), is always located at the side where the blueshifted strands connect with the filament crest, which strongly suggests a link between the velocity gradient over the filament crest and the immediate surrounding blueshifted strands. These observations point to a scenario where the strands are being accreted on the crest. As it is possible that this mass accretion is not exactly fixed on the filament axis, it can also deposit angular momentum inside the filament which could contribute to the C$^{18}$O(2-1) crest velocity field. This is schematically shown in Fig. \ref{sketchAccretion}.\\
Velocity gradients perpendicular to filaments have been found in other clouds as well, e.g.: DR21 \citep{Schneider2010}, IRDC 18223 \citep{Beuther2015}, Serpens \citep{Dhabal2018,Chen2020} and SDC13 \citep{Williams2018}. In these studies, the observed velocity gradients have also been found to be possible indications of mass accretion by the filament from inflowing lower-density gas. However, the interpretation is often complicated by e.g. the presence of multiple velocity components \citep[e.g][]{Schneider2010,Beuther2015,Dhabal2018}. In the southern map of Musca there is a hint of a second component in C$^{18}$O, but in the northern map the emission only shows a single velocity component which experiences the velocity gradient. This strongly suggests that the velocity gradient in Musca is indeed a mass accretion signature. 
%With the simple velocity structure of the Musca filament crest, we now had an unprecedented case to find a link between the nearby ambient cloud and the kinematics in the filament. 
%The presented correlation demonstrates that gradients perpendicular to the filament crest can be related to continuous mass inflow and that this might explain the transonic linewidth of fibers.
% It was already noted that the gradient in the fragmenting area appears a bit less clear. This could imply that it becomes more difficult to clearly establish the mass accretion signature when filamentary structures become more complex or evolved, e.g. through fragmentation, multiple fibers, longitudinal collapse,...

\subsection{A potential \HI cloud-cloud collision scenario to form the Chamaeleon-Musca complex}

We obtain a scenario where the dense Musca filament is continuously accreting mass from large scale inflow, which fits with observed indications of filament accretion shocks towards the Musca filament in Paper II. Here we study the large-scale kinematics of the Chamaeleon-Musca complex to constrain the mechanism that drives the continuous mass accretion of the Musca filament and thus is responsible for the formation of the Musca filament.

\subsubsection{A $50 - 100\,$pc coherent cloud complex in CO and \HI}

The results of the NANTEN CO survey of the Chamaeleon-Musca region in \cite{Mizuno2001} show the existence of a well-defined velocity ($-1$ to $6\,$km s$^{-1}$) and spatially coherent CO complex extending over roughly $20^\circ\times24^\circ$, i.e. $35-40\,$pc at 90 pc to $70-80\,$pc at 190 pc while the analysis of GAIA data towards the CO clouds Cha I, II, III and Musca indicates walls of extinction ranging from 90 to 190 pc along the line of sight. This suggests that the global CO gas and the well known clouds Cha I, II, III and Musca are parts of a single complex of size 50 to 100 pc in projection and along the line of sight. 

\begin{figure}
\includegraphics[width=\hsize]{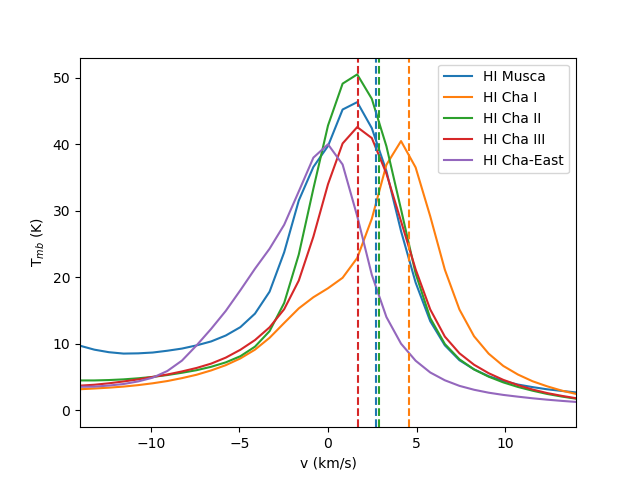}
\caption{\HI spectra averaged towards Cha I, Cha II, Cha III, the eastern region of the Chamaeleon cloud and Musca. The dashed vertical lines give the $^{12}$CO velocities of the dense regions Cha I, Cha II, Cha III and Musca \citep{Mizuno2001}. This shows that the dense gas traced by $^{12}$CO for all regions is redshifted compared to their \HI emission, identical to what is found for the filament crest of Musca when comparing $^{12}$CO and C$^{18}$O.}
\label{HIspec}
\end{figure}

\begin{figure*}
\begin{center}
\includegraphics[width=\hsize]{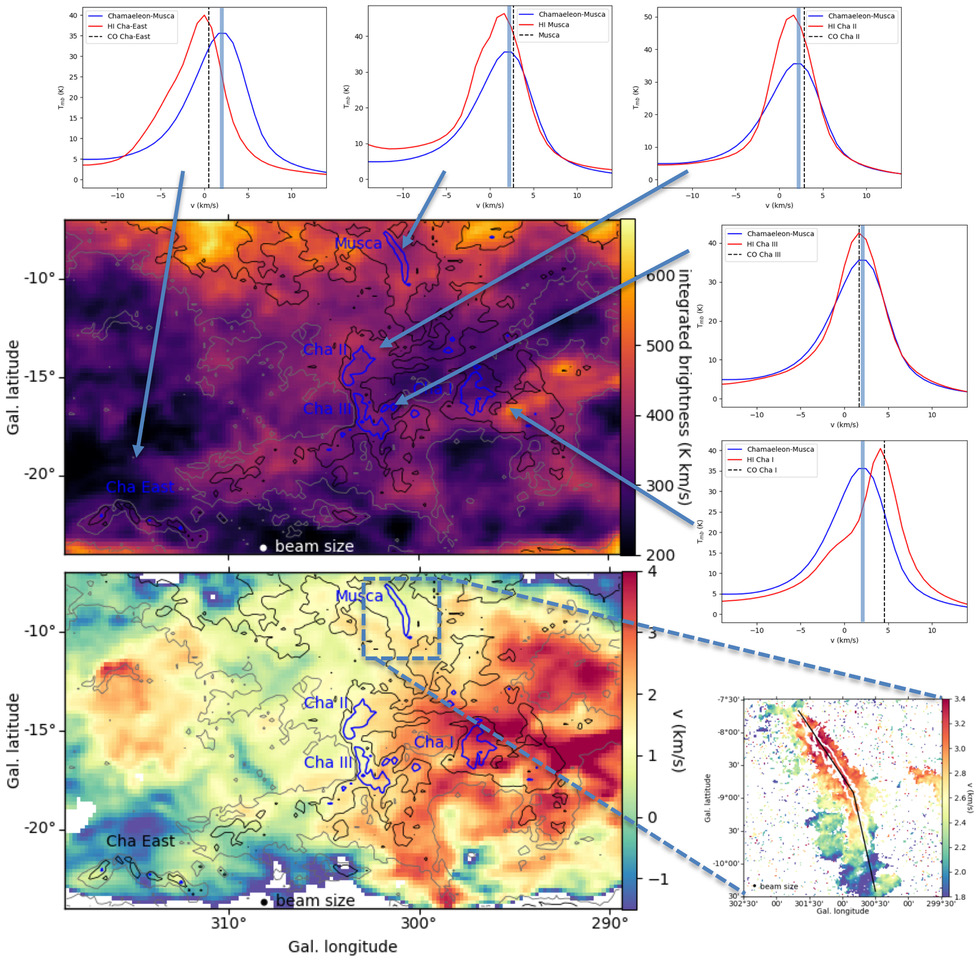}
\end{center}
\caption{Integrated \HI map (top) and the velocity field (below) of the Chamaeleon-Musca complex from fitting a single gaussian to the \HI spectra, with overplotted Planck brightness contours at 353 GHz. The blue contours highlight the dense Chamaeleon molecular clouds and the Musca filament, while the black and grey contours indicate the more extended continuum emission from Planck. 
Around the two maps, the average \HI spectrum of the full complex and its peak emission (blue) is compared with the local \HI spectrum (red) and velocity of the CO gas (black dashed line) \citep{Mizuno2001} from the selected dense region (Cha I, Cha II, Cha III, Cha East and Musca). In the lower right figure, the $^{12}$CO velocity field from the NANTEN2 data is shown.}
\label{HIvelMap}
\end{figure*}

To go one step further, we investigated the atomic hydrogen (\HI) 21 cm line from the Galactic All Sky Survey (GASS) with the Parkes telescope \citep{McClure2009,Kalberla2010,Kalberla2015} towards the Chamaeleon-Musca region. The data has a spectral resolution of 0.8 km s$^{-1}$ and a spatial resolution of 16$^{\prime}$ ($\sim$ 0.65 pc). This does not resolve CO clouds individually, but we find in Fig.~\ref{HIspec} that the \HI spectra have central velocities ranging from $-2$ to $5\,$km s$^{-1}$ for the CO clouds. In Fig.~\ref{HIvelMap} we show in the upper map the \HI integrated emission between -10 and 10 km s$^{-1}$ with contours of the Planck brightness at 353 GHz which indicates that there is a good spatial coincidence between \HI and extended dust emission for the Chamaeleon-Musca complex. 
Fitting a gaussian to the \HI spectra, we obtain a velocity field which is mapped in the lower map of Fig.~\ref{HIvelMap}. It shows that the whole complex displays a global velocity gradient from (slightly south) east at $-2\,$km s$^{-1}$ to (slightly north) west at $+5\,$km s$^{-1}$ which corresponds well to a similar $^{12}$CO east-west gradient along a filamentary feature $\sim 15^\circ$ long (i.e. 37 pc long in projection at 140 pc; \citealp{Mizuno2001}) which is also seen in Planck dust emission. %This velocity difference over the complex for \HI is similar to the $^{12}$CO velocity gradient found in \cite{Mizuno2001}, but slightly blueshifted with respect to $^{12}$CO. 
We note that the magnitude of this gradient is similar to what was found for giant molecular clouds in M33 and M51, for which it had been proposed that it would be related to galactic pro- and retrograde rotational motion \citep{Braine2018,Braine2020}. The global velocity field in the Chamaeleon-Musca complex might thus also be linked to some galactic rotation. However, here it would then be retrograde motion unlike clouds in M33 and M51 which are mostly found to be prograde.
While the velocity gradient in \HI over the region is roughly identical to that of CO, the \HI linewidths {of $\sim$ 5-6 km s$^{-1}$} are slightly larger than the linewidths in CO (2 to 2.8 km s$^{-1}$ in \cite{Mizuno2001}, as expected if the turbulent motions in \HI are driving the CO motions as proposed for instance in numerical simulations of cloud formation by \citet{Koyama2002}.\\\\

It is therefore clear that a coherent cloud complex exists connecting all known CO clouds of the region (Cha I, II, II and Musca) as well as a few additional more diffuse CO clouds such as Cha East and smaller CO clouds discussed for instance in \cite{Mizuno1998}.

\begin{figure*}
\begin{center}
\includegraphics[width=0.9\hsize]{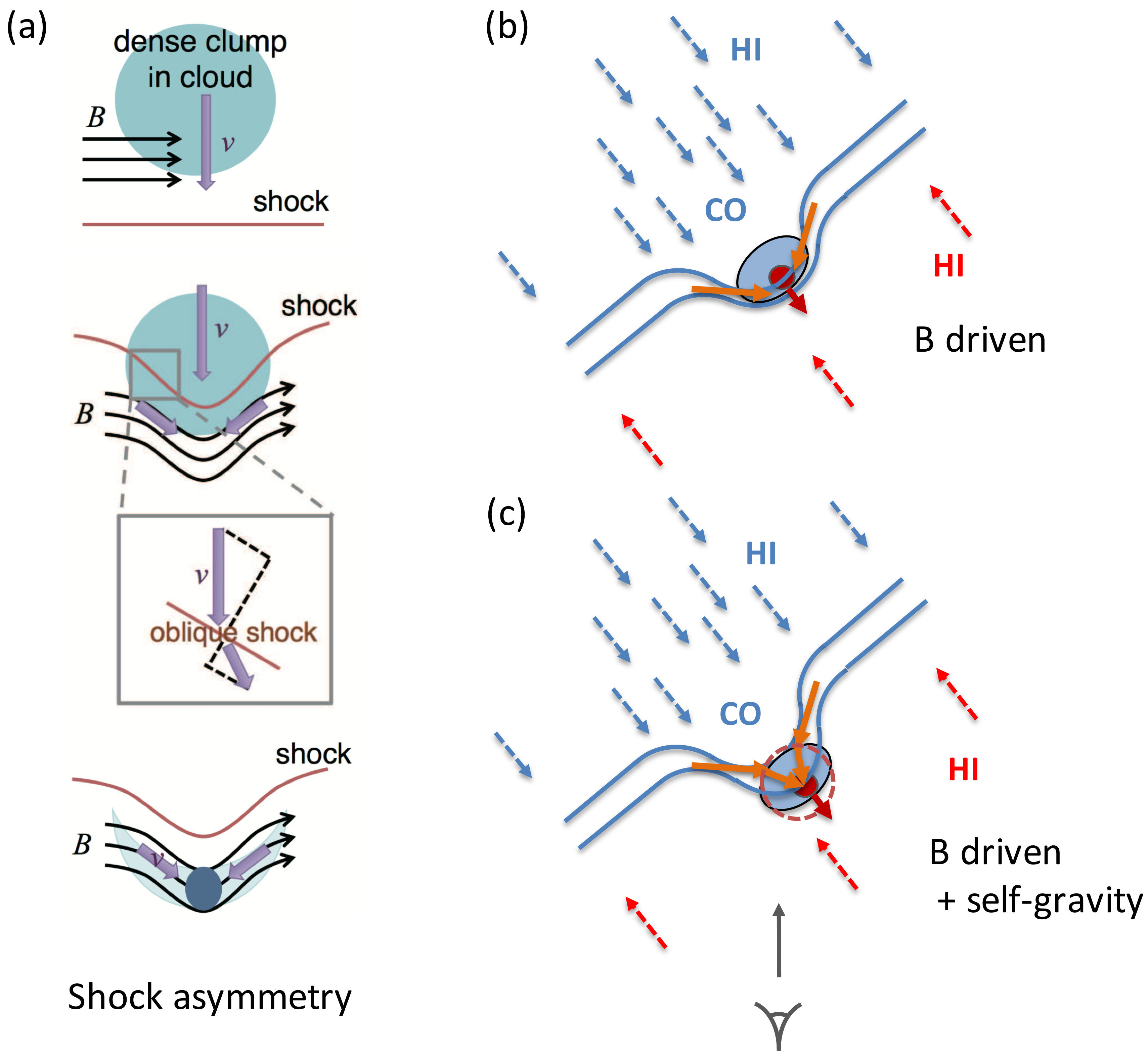}
\end{center}
\caption{Proposed sketch, observed from above the Musca filament, to explain the spatial and kinematic asymmetries of Musca originating from a \HI cloud-cloud collision event (blue and red dashed arrows) following the Inoue scenario (see panel (a) \cite{Inoue2018}). The asymmetry between CO and \HI could be due to the effect of %slowing down of the densest, CO structures as they accumulate momentum from the red-shifted \HI gas through the magnetic tension from large scales and behind the front shock.
 compression of a clump in a cloud-cloud collision. The resulting bent of the magnetic field (blue lines) drives the gas towards the apex along the magnetic field, leading to a concentration of the mostly blue-shifted but slowed down (i.e. becoming red-shifted compared to the local mostly blue-shifted \HI gas) CO gas  (B driven concentration, panel (b)). Later on, the self-gravity of the strong concentration of matter (filament crest) can curve the concentration flows (see panel (c)) continuing the accretion onto the crest while magnetic field slowly drifts through the crest. The crest would then be in equilibrium between an asymmetric accretion and the tension of magnetic field. %This would explain the velocity gradient through the mostly eastern CO surroundings of the Musca crest. 
 The progressive slowing down of the accretion flow, observed in the strands, from the bluest gas to the crest velocity could be partly due to magnetic pressure (the flows get less aligned with the magnetic field as self-gravity starts to take over) and perhaps some contribution from the turbulent pressure as it gets close to the crest (see Sect.~\ref{sec: GravityShimajiri}).   
%A sideway view of this pattern would put the most red-shifted CO gas more or less coincident with the crest of the filament in the north-western parts of the more blue-shifted strands/ambient CO gas such as observed. 
Seen from the side, the magnetic field would naturally appear perpendicular to the crest of the filament as observed.}
\label{sketchInoue}
\end{figure*}

\subsubsection{Dense gas mostly red-shifted in Chamaeleon/Musca}

Studying the CO isotopologues towards Musca, we found that the dense filament crest, traced by C$^{18}$O, is at redshifted velocities compared to the molecular ambient cloud traced by $^{12}$CO (Sect.~\ref{sec: NantenSection}). With \HI, it is possible to study the kinematics of the even lower density and larger scale ambient gas. Inspecting the average \HI spectrum towards the Musca cloud in Fig.~\ref{HIspec}, we found that the NANTEN2 $^{12}$CO emission also appears redshifted compared to \HI in the same way as C$^{18}$O compared to $^{12}$CO at smaller scale in  Fig.~\ref{averageSpecNewComp}. This demonstrates that denser gas in the Musca cloud is systematically redshifted from low density, large scale atomic and molecular gas down to the Musca strands and filament crest.

In Fig.~\ref{HIvelMap}, we plot the individual \HI spectra (in red) of the Chamaeleon/Musca clouds (Cha I, Cha II, Cha III, Cha East and Musca) together with their $^{12}$CO velocities reported in \cite{Mizuno2001} (dashed lines) in comparison with the globally averaged \HI spectrum (in blue). We see that while the individual \HI spectra are equally present in the blue and red-shifted parts of the average \HI spectrum, the CO gas tends to be mostly in the red-shifted part and is basically always red-shifted compared to the individual \HI spectra towards each regions. This suggests that the velocity asymmetry of the dense gas, being mostly red-shifted, seen towards Musca is a general trend for the whole Chamaeleon-Musca cloud complex.
%Looking into the \HI spectra of the dense Chamaeleon clouds (Cha I, Cha II and Cha III), and comparing it with their $^{12}$CO velocity reported in \cite{Mizuno2001} (see sub-panels in the top and right parts of Fig.~\ref{HIvelMap}), shows that also for these dense clouds the denser $^{12}$CO emission is systematically towards the redshifted velocities of the \HI emission. Indeed while the average individual \HI spectra are equally present in the blue and red-shifted parts of the \HI emission, the CO gas tends to be mostly in the red-shifted part and is always slightly red-shifted compared to the average emission in \HI towards each regions such as noted above for the Musca region. This suggests that the asymmetry of the velocity of the dense gas seen towards Musca is a general trend of the whole Musca-Chamaeleon cloud complex.
Combining this with the identical redshifted asymmetry at small scales for Musca suggests a dynamical coherence in the complex, i.e. similar kinematic asymmetry, from the \HI cloud with a size of $\sim 50-100\,$pc down to $\sim0.1\,$pc around the Musca filament crest.
%One has to take into account that we only trace the velocity along the line of sight, but the dynamic asymmetry is found for basically all molecular $^{12}$CO gas of the Chamaeleon-Musca complex.
This puts forward a scenario where the mass accretion of the dense (star forming) gas, which was in particular proposed based on the velocity gradients over the filament crest, in the complex is directly related to this large scale, and asymmetric \HI/CO kinematics of the full complex. %We note that this conclusion fits with the similarity of turbulent properties in Cha I, Cha II and Cha III observed by \cite{AlvesDeOliveira2014}.

%We note in addition that inspecting the average \HI spectra towards the dense regions of the Chamaeleon-Musca complex, see Fig. \ref{HIspec}, shows that the spectra towards Musca and Cha I indicate the presence of two components. This suggests the presence of at least two velocity components leading to a shear or colliding flow of the blue- and redshifted \HI components in the Chamaeleon-Musca complex. However, in Cha II and Cha III the presence of multiple components is not immediately clear. 

We conclude that the local kinematic asymmetry observed towards Musca is observed for the whole complex and individually for each CO cloud of the complex. This suggests that the region corresponds to a single event of cloud formation. {We argue in the following subsection that a \HI cloud-cloud collision event can explain the kinematic asymmetry, which can fit with the observed indications of more than one velocity component in several \HI spectra.}

\subsection{Asymmetric inflow guided by the magnetic field in a \HI cloud-cloud collision}
\label{sec: InflowGuidedMagField}

%\subsubsection{An asymmetry collision to explain global asymmetries}

The asymmetry in velocity between low and high density gas at all scales is difficult to explain with a classical view of an isotropic injection of turbulence at large scales. In this view, the velocity streams would be on average equally blue and red-shifted. In contrast, any scenario based on a (\HI) cloud-cloud collision can easily introduce asymmetries if the two clouds in collision have different initial properties and substructure, which is actually expected and natural. %If the blue-shifted \HI cloud is less massive than the red-shifted one, the densest regions (CO gas) would be mostly made of original red-shifted cloud material. The red-shifted \HI cloud could also contain more concentrated structures than the blue-shifted one leading to a collision which would enhance preferentially these red-shifted structures to form the observed dense molecular structures.
%\subsubsection{Spatial and kinetic asymmetry as a hint for Inoue's scenario for Musca}
%It is generally considered that inflow towards the dense filaments is guided by the magnetic field \citep[e.g.][]{Banerjee2009,Soler2013,Inoue2013,Palmeirim2013,Cox2016,Fogerty2017}. It does however remain difficult to get conclusive some evidences of this. 
In the case of a (\HI) cloud-cloud collision with some primordial density substructure, numerical simulations show that such a collision is prone to concentrate the pre-existent structures \citep[e.g.][]{Inoue2012,Inoue2013,Inoue2018,Iwasaki2019}. This concentration of matter can originate from a locally curved magnetic field leading to dense structures perpendicular to the magnetic field. This would fit with the generally accepted view that inflows towards the dense filaments are guided by the magnetic field \citep[e.g.][]{Banerjee2009,Soler2013}, possibly in the form of striations \citep{Palmeirim2013,Cox2016}. This (Inoue) mechanism is based on the fact that when a shock wave impacts a slightly denser structure, the local magnetic field is bent around the structure and then channels the streams of gas (through an oblique shock; see Fig.~\ref{sketchInoue}a) towards the apex of the bent which is then rapidly the densest part of the original structure \citep{Inoue2013,Vaidya2013}. 

\begin{figure}
    \centering
    \includegraphics[width=\hsize]{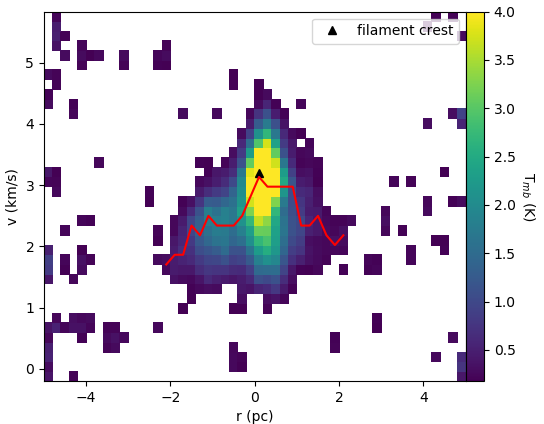}
    \caption{PV diagram as a function of the distance from the center of the Musca filament for the $^{12}$CO(1-0) emission from the NANTEN2 data. The red line follows the velocity with maximal brightness as a function of the distance from the filament crest. The PV diagram shows the expected V-shape from the Inoue scenario with respect to the velocity of the dense filament crest \citep{Hacar2016} which is at the redshifted apex of the V-shape. %{\bf Right}: The PV diagram towards the Musca cloud of the \HI emission from the GASS survey, with overplotted the $^{12}$CO(1-0) velocity structure. It is observed that the \HI emission is less affected by the presence of the Musca cloud than CO, indicating that the \HI emission indeed traces the kinematics of the Chamaeleon-Musca complex.
    }
    \label{PVdiagrams}
\end{figure}

Interestingly enough such a bent of the magnetic field by a shock in the Inoue mechanism naturally predicts a spatial and kinematic asymmetry of the dense structure with the densest region, the filament crest, being concentrated on the front side of the original structure. Fig.~\ref{sketchInoue} illustrates how such a scenario could explain the described kinematical asymmetry of Musca as well as the indicated spatial asymmetry in Sec. \ref{sec: NantenSection}. The Musca filament/crest would trace this front side of the original structure which has been concentrated by favoured channelling parallel to the bent magnetic field (see the sketch in the right panels of Fig.~\ref{sketchInoue}). This dense structure is also the most slowed down gas by the opposite stream of gas from the red-shifted \HI cloud %as it drags red-shifted gas along an extended region through the magnetic field anchored in the crest 
(see panel (b) of Fig.~\ref{sketchInoue}). It would explain why all dense CO structures in the region appear red-shifted as they would be the most sensitive to the dragging force of the opposite flow of gas from the colliding red-shifted \HI cloud. If globally \HI is dominated by the blue-shifted \HI gas, this would explain the tendency to have the \HI spectrum slightly blue-shifted compared to the CO gas of all molecular clouds in the Cha/Musca region.
%In order to explain in addition the spatial asymmetry at small scale observed for Musca, we may consider a particular case of cloud-cloud collision, namely the simulation driven view proposed by Inoue+ where pre-existent structures can be concentrated and may accumulate material thanks to a channelling of the magnetic field behind the shock front generated by the collision. 
In this scenario the magnetic field would channel the pre-existent mostly blue-shifted structure and its immediate surroundings, while slightly slowing down and becoming more red-shifted and therefore allowing for more accretion of faster blue-shifted gas from behind. The orientation of this organised inflow by two intersecting sheet-like structures could then explain why most CO emission is observed east of the filament. This scenario has the advantage to explain both the spatial and velocity asymmetries, and to explain the fact that the magnetic field is perpendicular to the filament/crest. \\\\
In particular, this bending of the magnetic field would lead to an observable signature in the PV diagram of the Musca cloud. This signature is displayed in Fig. \ref{sketchInoue} and in Fig. 16 from \cite{Arzoumanian2018}. Specifically, the kinematic signature consists of a $'$V-shape$'$ for the inflowing molecular cloud with the densest gas at the apex of this V-shape. A V-shape in the PV diagram was reported by \cite{Arzoumanian2018} for a filament in the Taurus molecular cloud. Fig. \ref{PVdiagrams} shows the PV diagram of the Musca cloud from the NANTEN2 data. This PV diagram shows exactly such a V-shape, with the C$^{18}$O velocity of the filament crest at the apex of this V-shape. %When comparing the PV diagram of the NANTEN 2 data with the results for the \HI data in Fig. \ref{PVdiagrams}, it is observed that the \HI emission displays a different behaviour than the CO gas. This indicates that the \HI emission indeed traces the larger scale kinematics associated with the Chamaeleon-Musca complex. Due to the relatively low angular and spectral resolution of the \HI data for a detailed study of the Musca cloud, it is currently not possible to study the presence of self-absorption related to the formation of the Musca cloud.

\subsection{Inhomogeneities in the inflow driven by local gravity or large-scale dynamics}

\begin{figure}
\includegraphics[width=\hsize]{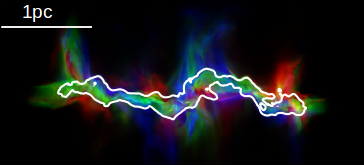}
\caption{RGB image using $^{12}$CO(1-0) synthetic observations (red: 0.5 to 2 km s$^{-1}$ \& blue: -2 to -0.5 km s$^{-1}$) and $^{13}$CO(1-0) synthetic observations (green: -0.5 to 0.5 km s$^{-1}$) of the simulations in \cite{Clarke2018}. The white contour encloses the region with integrated C$^{18}$O(1-0) brightness $>$ 1 K km s$^{-1}$. This shows that close to the filament the direction of the velocity gradient can change, similar to what is observed in Musca, as a result of turbulent motions.}
\label{rgbSeamus}
\end{figure}

While the Inoue scenario may well explain the observed local asymmetries towards the Musca filament, the change of direction for the velocity gradient along the filament discussed in Sect.~\ref{sec: massAccretion} might be more difficult to explain with this scenario alone. We note however that numerical simulations such as the ones presented in \cite{Clarke2017,Clarke2018} reproduce very well this behaviour (see also simulations presented in \citealp{Schneider2010} to explain similar features in the DR21 filament/ridge). These simulations consist of a cylindrical converging flow which produces an accreting dense filament. Turbulence in the flow leads to inhomogeneities in the accretion and substructures in the filament with accretion flows which are not always landing on the filament from the same side of the crest. The simulation includes hydrodynamics, self-gravity, a weak FUV-field (1.7 G$_{0}$), and heating and cooling coupled with non-equilibrium chemistry. This allows for the self-consistent formation of CO and thus the production of synthetic CO observations as shown in Fig.~\ref{rgbSeamus}, which present an RGB image from the simulation that highlights the locations of the blue- and redshifted mass reservoir around the dense filament. These synthetic observations indicate that there is no clearly separated blue- and redshifted mass reservoir close to the filament crest, but rather that the location of the dominant inflowing mass reservoir alternates to each side of the filament as a consequence of the turbulence in the inflowing mass reservoir \citep[][Fig. \ref{rgbSeamus}]{Clarke2017}. 
The simulations demonstrate that in a cylindrical area close to the filament there can be large local position variations of the converging flows towards the filament. In Musca this relatively cylindrical area has a size of the order of 0.4 pc, covering the filament crest and strands. %This cylindrical area might be the intersection of two parsec scale flows, as the NANTEN data show indications that the Musca filament splits the velocity field of the ambient cloud on large scales in a blue- and (slightly) redshifted region similar to B211/3 \citep{Palmeirim2013}, see Figs. \ref{largeMapNANTEN} and \ref{fitsNANTEN}.

We conclude that these inhomogeneities in the inflow can be expected at least in a non-magnetic case (the above discussed simulations do not include magnetic field).
It is not clear though what is the main driver of these local variations of the converging flows in Musca. Including the possible role of magnetic field, it can either originate purely from the original kinematic fluctuations from large scales, forcing the magnetic field guidance or could be partly due to self-gravity close to the crest which could decouple the gas inflow from the guidance of magnetic field allowing for these variations along the filament.\\

%On the other hand these simulations do not include magnetic field and start from peculiar initial conditions for which the initial cloud is collapsing and is already elongated . Comparing these simulations with those of Inoue's group with magnetic field, this might indicate that the large-scale concentration of matter could lead to the spatial asymmetries, and be due to the magnetic field bending after a shock front of a cloud-cloud collision and that the obtained elongated cloud may then become self-gravitating with a collapse at the scale of a few 0.1 pc leading to some level of variations in the geometry of the accretion flow along the filament.

\subsection{A two step scenario: B driven followed by a gravity driven and  B regulated accretion}
\label{sec: GravityShimajiri}

%At the Musca filament crest, we found indications of mass accretion that might be related to a dynamic coherence from the \HI cloud down to 0.1 pc. A next step is to understand the physical process responsible for this behaviour.

As shown in the previous section thanks to numerical simulations, self-gravity of the filamentary structure may explain the changing direction of the velocity gradients along the crest of the filament. We may then have a situation as shown in panel (c) of Fig.~\ref{sketchInoue} where there is a transition from the B (magnetic field) driven concentration to a gravity driven accretion regime close to the crest. This may correspond to the region of the strands where we observe a smooth $^{13}$CO velocity gradient from 3.1 to 3.5 km s$^{-1}$ and from 2.7 to 3.1 km s$^{-1}$ for the northern and southern APEX maps, respectively. The progressive slowing down of the blue-shifted gas to the velocity of the crest could then be due to the need to cross some magnetic field lines to reach the crest. Indeed, close to the crest where the self-gravity of the crest is maximal, the magnetic field should slowly drift through the filament crest in the Inoue scenario (Fig.~\ref{sketchInoue}) as it is dragged by the larger scale pressure from the red-shifted gas. So we would have a scenario where the gas is accelerated by gravity but slowed down by magnetic pressure to reach the up-stream crest.

Alternatively in \citet{Shimajiri2018} for the B211/3 filament in Taurus, gravity was claimed to accelerate gas from large to small scales, and with a slowing down of velocities close to the crest being due to a large turbulent pressure (up to an order of magnitude larger than the thermal pressure). To obtain a fit to the data, they had to assume that the whole CO linewidth would represent an effective turbulent pressure.
In Musca in the APEX and NANTEN2 maps we see that the global velocity dispersion appears to be dominated by bulk motions with velocity gradients, and not by isotropic turbulence. We easily expect that at higher spatial resolution an even larger fraction of the linewidth is due to bulk motions rather than to isotropic turbulence. Also we note that 
the velocity dispersion (expressed in $\sigma$) is much smaller in Musca compared to B211/3. Instead of 0.9 km s$^{-1}$ for B211/3, we obtain a maximal $\sigma = 0.4$ to $0.5\,$km s$^{-1}$ over the whole region for Musca in the NANTEN2 $^{12}$CO data (e.g. Fig.~\ref{fitsNANTEN}), and even down to $\sigma \sim 0.2\,$km s$^{-1}$ close to the crest in the $^{13}$CO APEX data. The possible deceleration thanks to a possible contribution of the turbulent motions to pressure in the immediate surroundings of the crest is therefore reduced in Musca compared to B211. Since the gravitational acceleration in Musca is also reduced (lower mass filament) it is not possible to firmly reject some effect of the turbulent pressure effect, but we point out here that our scenario can explain the observed deceleration without heavily relying on an hypothetical full conversion of CO linewidth into effective sound speed such as in \citet{Shimajiri2018} for B211/3. Altogether we may actually have a mixture of magnetic and turbulent pressure to explain the deceleration of the accretion flow onto the crest.

We finally note that our scenario might also explain the geometry and velocity field of the B211/3 filament. This filament shows an asymmetry between the blue- and redshifted sides of the accretion flow, and {also displays a V-like shape in the PV diagram} \citep{Shimajiri2018}. The proposed angle of 130$^{\circ}$ between the two sheets in \cite{Shimajiri2018} could actually trace the bent of the magnetic field in the Inoue scenario.

%\subsubsection{Accretion driven fiber linewidth?}
%The velocity gradients over the crest cover a velocity range of $\sim$ 0.15-0.2 km s$^{-1}$, which corresponds well to the non-thermal velocity dispersion found in the Musca filament crest {\bf in this study as well as in \cite{Hacar2016}}. This suggests that deposited angular momentum from mass accretion on the crest provides the dominant contribution to the non-thermal motion in the transonic Musca filament crest. In \cite{Traficante2020} it was recently also put forward that mass inflow towards massive clumps can significantly affect the observed linewidth.\\ 

~%Taking a more universal look at fibers and low-mass filaments below the critical line mass\footnote{The critical line mass is defined by 2c$_{s}^{2}$/G \citep{Ostriker1964}, and is the maximal line mass for which a non-magnetised isothermal filament is stable to perturbations} ($\lesssim$ 16 M$_{\odot}$ pc$^{-1}$), it is quite generally found that they have similar transonic linewidths \citep{Arzoumanian2013,Hacar2018}. This could be an indication that the transonic linewidth in low-mass fibers and filaments can be the result of angular momentum that is deposited as a result of mass accretion.

\subsection{The physical scale of dominant self-gravity}
\label{sec: GravityScale}

%Gravity is a long distance force that leads to collapse if there is no other force to prevent gravity from dominating the kinematics. 
Here we compare the typical observed relative motions between the crest and the surrounding gas with the expected self-gravity velocities to discuss at which scale gravity may be responsible for the observed relative motions. 
%The increasing role of self-gravity close to crest is also supported by the magnitude of the velocity gradients. 
In \cite{Chen2020} it is proposed to use the non-dimensional parameter C$_{\nu}$ to differentiate between gravity-driven mass inflow and other sources of motions (shock compression in \citealp{Chen2020}), with C$_{\nu}$ expressed as
\begin{equation}
    C_{\nu} = \frac{\Delta v_{h}^{2}}{G\:M(r)/L}
\end{equation}
where $\Delta v_{h}$ is half of the velocity difference over the filament and M(r)/L the mass per unit length at the distance r from the center of the crest. 

On large scales, the velocity difference between the filament and the CO gas reaches values of the order of 1 km s$^{-1}$ at both sides of the filament crest, see Fig. \ref{PVdiagrams}. With a mass per unit length of 15.6$\,$M$_\odot$ pc$^{-1}$  (22.3$\,$M$_\odot$ pc$^{-1}$ corrected to d = 140 pc) for the Musca filament, this corresponds to C$_{\nu}$ = 15. At the pc scale, the kinematics are thus not dominated by the self-gravity of the filament. In other words the actual crossing time is clearly smaller than the free-fall time from the self-gravity of the filament. At the 0.4 pc scale of the strands, we still have a velocity difference of $\sim 0.4\,$km s$^{-1}$ leading to C$_{\nu}$ = 2.4. This suggests that self-gravity may start to play a role but is not dominant yet. At the scale of the filament crest with a velocity difference of 0.2 km s$^{-1}$ we obtain C$_{\nu}$ = 1.1. We thus find that gravity may indeed take over from the large scale motions at sub-pc scales, reinforcing our proposed scenario in two steps with a B-driven followed by a gravity driven accretion when matter is reaching the crest of the filament at sub-pc scale.

When we apply the C$_{\nu}$ criterion to study the importance of self-gravity for the B211/3 filament, we similarly found large values of C$_{\nu}$ at the pc scale, and at the sub-pc scale close to the crest that C$_{\nu}$ could be approaching a value of 1. The velocities of the large scale flows in B211/3 are found to be typically 2-3 times larger than in Musca at the same distances from the crest while the $M/L$ value is roughly 3 times larger (54$\,$M$_\odot$ pc$^{-1}$). C$_{\nu}$ is therefore found to be typically 2-3 times larger in B211/3 at large scales.% It is pointing to a least importance of self-gravity in B211/3 compared to Musca. %The larger effect of large scale motions to drive accretion flows in B211/3 may explain why several fibers with slightly different velocities have emerged in B211/3 in contrast to Musca. The velocity differences between the fibers are actually fitting well with the expected excess of kinetic energy which had to be transfered in bulk motions between fibers to let self-gravity takes over at small scales to increase even more densities of the crest(s) of fibers. 

\subsection{Future star formation in the Musca filament}
In this paper we proposed that a large scale colliding \HI flow, forming the Chamaeleon-Musca complex, could be at the origin of the proposed mass accretion on the Musca filament crest. Here, we investigate whether such continuous mass accretion can eventually lead to star formation in the Musca filament.\\\\
In the NANTEN2 data, most molecular gas traced by $^{12}$CO is present in the filament and in a structure east of the filament which has a relative velocity of -0.5 to -1 km s$^{-1}$ compared to the velocity of the filament. Using simple assumptions, we now estimate a time scale for accretion of this gas on the filament. We assume a projected inflow velocity in the plane of the sky of 0.7 km s$^{-1}$ towards the filament, while the distance perpendicular to the crest of this $^{12}$CO emission is generally of the order of 0.5 to 1 pc. It would thus require roughly 0.7 - 1.4 Myr for the nearby ambient cloud within this radius to be accreted on the filament/strands. Then there are the strands at a velocity of -0.4 to -0.5 km s$^{-1}$ along the line of sight compared to the filament crest. These strands are present from the edge of the filament crest up to a distance of $\sim$ 0.4 pc. At this velocity, the accretion of these strands on the filament crest would take close to 1 Myr. The filament crest has a velocity gradient, with a maximal velocity difference of 0.15 to 0.2 km s$^{-1}$. This results in a similar crossing timescale for the filament crest of 0.7 Myr. However, one should note that this velocity gradient is not necessarily related to dispersion of the filament, and that the increasing role of gravity will confine the filament such that this timescale is a lower limit.\\\\
Similar to \cite{Palmeirim2013} we can estimate the mass accretion rate per unit length on the filament at R = 0.4 pc. For this we need a density estimate at R = 0.4 pc. Here we use the same approach as \cite{Palmeirim2013} by extracting the density from the fitted Plummer model to the averaged Musca filament \citep{Cox2016}. This results in a mass accretion estimate of 14 M$_{\odot}$ pc$^{-1}$ Myr$^{-1}$. With this mass accretion rate it would roughly take 1 Myr to accrete the amount of mass already present in the Musca filament. Based on the estimated mass accretion rates and time scales, we find that strands can be provided with sufficient mass from the more extend ambient cloud during the timescale of their accretion on the filament crest. This allows for a continuous mass accretion on the Musca filament.\\\\
Several theoretical studies have investigated the collapse timescales of non spherical structures \citep[e.g.][]{Burkert2004,Toala2012,Pon2012b,Clarke2015}. To estimate the longitudinal collapse timescale of the Musca filament, we use the formula from \cite{Clarke2015}:
\begin{equation}
    t_{coll} = (0.49 + 0.26A_{0})(Gn_{H_{2}})^{-\frac{1}{2}}
\end{equation}
Where A$_{0}$ is the initial aspect ratio, which is 54 using a half length of 3 pc and a radius of 0.056 pc for Musca \citep[][corrected to d = 140 pc]{Cox2016}, G is the gravitational constant, and for n$_{H_{2}}$, the molecular hydrogen gas density, we use 10$^{4}$ cm$^{-3}$. This results in a timescale of 9.2 Myr, which is roughly an order of magnitude larger than the estimated accretion time scale for the large scale inflowing mass reservoir. This lifetime for longitudinal collapse is a lower limit because of possible magnetic field support.\\
%If the Musca filament crest would follow Larson's relation \citep{Larson1981}, one can also estimate a lifetime for the Musca filament from the velocity dispersion:
%\begin{equation}
%\sigma (\text{km s}^{-1}) = 1.1 \text{km s}^{-1} (\text{L/pc})^{0.38}
%\label{larsonEquation}
%\end{equation}
%L = 6 pc then gives $\Delta$v = 2.35$\cdot \sigma$ = 5.2 km s$^{-1}$. The resulting timescale for dispersal of the Musca filament would be of the order of L/$\Delta$v = 1.1 Myr. However,
%Furthermore, it was demonstrated by \cite{Hacar2016} that the transonic Musca filament is decoupled from the Larson regime.\\\\
We thus find a scenario where the Musca filament crest is a long-lived filament because it is a coherent structure with sufficient continuous mass accretion which is driven by the bending of the magnetic field due to the large scale colliding \HI flow that forms the Chamaeleon-Musca complex. As this provides a sufficient amount of mass for Musca to become supercritical before it is dispersed, gravity starts playing a more important role in further confining the filament. The future continuous mass accretion, which might be increasingly driven by gravity, could further increase the density in the filament. This can lead to further fragmentation of the Musca filament, which has started in several regions of the filament \citep{Kainulainen2016}, to form pre-stellar and protostellar cores in a dynamic filamentary structure.

\section{Conclusion}
We have presented APEX observations of CO(2-1) isotopologues towards the Musca filament crest and the strands. This data was complemented with NANTEN2 observations of $^{12}$CO(1-0) covering the full Musca cloud. We find that C$^{18}$O traces the filament crest, $^{13}$CO the strands and $^{12}$CO the more diffuse ambient molecular cloud. The main results of this study can be summarised as follows:\\\\
- GAIA star reddening data suggests a distance of 140 pc for Musca, but we note that there might already be some reddening at $\sim$ 100 pc.\\\\
- Modelling the CO lines with non-LTE line radiative transfer favours a scenario where Musca crest is a compact filament with a central density n$_{H_{2}}$ $\sim$ 10$^{4}$ cm$^{-3}$ at T$_{\text{K}}$ $\sim$ 10 K, and where the strands are an independent feature, consisting of dense gas (with n$_{H_{2}}$ $\sim$ 10$^{3}$-3$\cdot$10$^{3}$ cm$^{-3}$ at T$_{\text{K}}$ $\sim$ 15 K) that is connected to the filament crest.\\\\
- We report a sharp increase of the [$^{13}$CO]/[C$^{18}$O] abundance ratio by roughly an order of magnitude over a small distance ($<$ 0.2 pc) at A$_{V} <$ 3. This occurs in a weak ambient FUV field ($<$ 1 G$_{0}$) and indicates C$^{18}$O is a limited column density tracer.\\\\
%s- At the filament crest we find [NH$_{3}$]/[H$_{2}$] $<$ 10$^{-9}$.\\\\
- We confirm that the filament crest is a velocity-coherent structure, and also demonstrate that there are transverse velocity gradients over the velocity-coherent filament crest with a magnitude similar to the transonic linewidth of the filament crest.\\\\
- The ambient cloud contains a significant amount of blueshifted gas along the line of sight with respect to the filament crest, while there is a lack of redshifted gas with respect to the filament crest.\\\\% This suggests an asymmetric ambient cloud for the Musca filament.\\\\
- We observe a link between the transverse velocity gradient over the filament crest and the location of the blueshifted strands, indicating that the strands are accreted on the filament crest and that the velocity gradient is a signature of this accretion which possibly also deposits some angular momentum in the crest.\\\\
%- The magnitude of the velocity gradient over the filament crest indicates that the transonic linewidth of the Musca filament crest {\bf can be the result of organised motion due to continuous mass accretion, with possibly a contribution of angular momentum deposition.}\\\\
%- The velocity gradient over the crest has an angle $\ge$ 30$^{o}$ relative to the large scale Planck magnetic field at both observed locations, suggesting that mass inflow is not perfectly aligned with the magnetic field close to the crest and/or that there is velocity reorientation during accretion on the filament crest. {\bf REMOVE THIS?}\\\\
- We find a kinematically coherent asymmetry from the $\sim$ 50 pc \HI cloud down to the Musca filament crest as well as an asymmetric column density profile for the Musca cloud, indicating that the large scale evolution of the Chamaeleon-Musca complex is directly related to the formation of the Musca filament and that the asymmetry contains essential information on the physical process responsible for dense gas formation in the region.\\\\
%- In the PV diagram of the Musca cloud, we find a V-shape with the velocity of the filament crest at the redshifted apex of this V-shape.\\\\
- The PV diagram of the Musca cloud traces a V-shape with the filament crest located at the redshifted apex of this V-shape.\\\\
Combining all data, we propose that the Musca filament crest is a long-lived dynamic filament because it is a coherent structure that is continuously replenished by inflowing gas. This mass accretion is driven by the colliding \HI flow that forms the Chamaeleon-Musca complex and can eventually lead to the formation of protostellar cores embedded in this dynamic filament. %It is not straightforward, at the moment, to explain the evolution of the Chamaeleon-Musca complex as a result of large scale gravitational collapse. 
This colliding flow in Chamaeleon-Musca is the result of a magnetised low-velocity \HI cloud-cloud collision that produces the observed asymmetric accretion scenario, seen as a V-shape in the PV diagram, driven by the bending of the magnetic field. 

\begin{acknowledgements}
    The authors thank the anonymous referee for providing insightful input that improved the clarity and quality of this paper.  L.B., N.S., R.S., and S.B. acknowledge support by the french ANR and the german DFG through 
the project "GENESIS" (ANR-16-CE92-0035-01/DFG1591/2-1). L.B. also acknowledges support from the R\'egion Nouvelle-Aquitaine. N.S. acknowledges support from the BMBF, Projekt Number 50OR1714 (MOBS-MOdellierung von Beobachtungsdaten SOFIA). N.S. and R.S.are supported by the German Deutsche Forschungsgemeinschaft, DFG project number SFB 956. We thank F. Wyrowski for providing information to correct for the frequency shift in the APEX FLASH observations. We thank F. Boulanger for input on the production of the Planck polarisation maps. This research made use of Astropy,\footnote{\url{http://www.astropy.org}} a community-developed core Python package for Astronomy \citep{Astropy2013,Astropy2018}. Based on observations obtained with Planck (http://www.esa.int/Planck), an ESA science mission with instruments and contributions directly funded by ESA Member States, NASA, and Canada. This publication is based on data acquired with the Atacama Pathfinder Experiment (APEX) under programme IDs 0100.C-0825(A), 0101.F-9511(A) and 0102.F-9503(A). This research has made use of data from the Herschel Gould Belt survey  
project (\url{http://gouldbelt-herschel.cea.fr}). The HGBS is a Herschel Key Project jointly 
carried out by SPIRE Specialist Astronomy Group 3 (SAG3), scientists of several 
institutes in the PACS Consortium (CEA Saclay, INAF-IAPS Rome and INAF-Arcetri, 
KU Leuven, MPIA Heidelberg), and scientists of the Herschel Science Center (HSC).
\end{acknowledgements}

% WARNING
%-------------------------------------------------------------------
% Please note that we have included the references to the file aa.dem in
% order to compile it, but we ask you to:
%
% - use BibTeX with the regular commands:
   \bibliographystyle{aa} % style aa.bst
   \bibliography{kinematics.bib} % your references Yourfile.bib
%
% - join the .bib files when you upload your source files
%-------------------------------------------------------------------

\begin{appendix}
\section{Distance of the dense Musca-Chamaeleon clouds}
\label{app: distanceAppendix}
\begin{figure}
    \centering
    \includegraphics[width=\hsize]{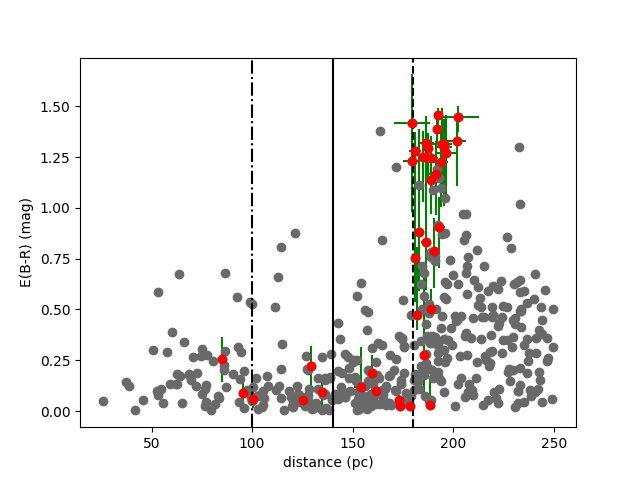}
    \includegraphics[width=\hsize]{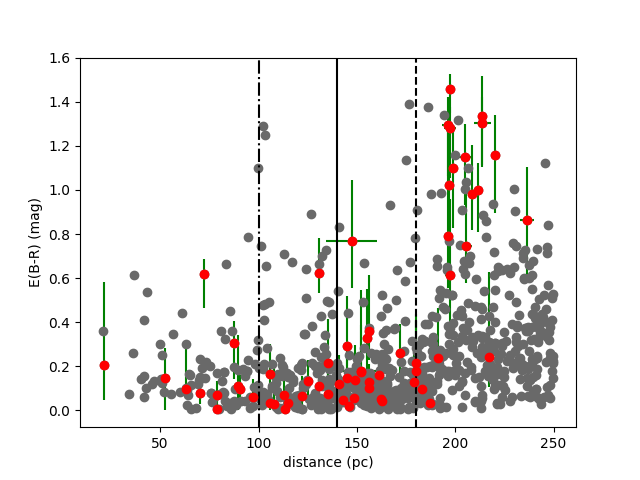}
    \caption{{\bf Top}: Reddening as a function of distance for stars in the area covered by Cha I. As in Fig. \ref{distMusca}, the red points are located at the high density region of the cloud. This shows that the reddening experiences a strong jump around 180 pc as was found in \cite{Zucker2019}. The other two vertical lines in the plots indicate a distance of 100 and 140 pc. {\bf Bottom}: The same for the region Cha II and Cha III together, showing a reddening jump at 190 pc.}
    \label{distancesCham}
\end{figure}

With the recent GAIA DR2 \citep{GaiaDR2} data release, a vast amount of information on nearby stars in the Galaxy has become available. This makes it possible to estimate the distance of interstellar clouds by studying the reddening they cause on background stars of these clouds \citep[e.g.][]{Zucker2019,Yan2019}. To estimate the distance of Musca, a region in the sky centered on the Musca filament with a radius of 80$^{\prime}$ ($\sim$ 3 pc at a distance of 150 pc) was selected in the GAIA catalogue, see Fig. \ref{PlanckMapChamaeleon}. In this region, all stars with a calculated reddening \citep{Andrae2018} and a distance $<$ 250 pc were selected. This results in a total of 223 stars found that can be used to estimate the distance of Musca. Because there can be foreground reddening we only select the stars that are located in the Planck 353 GHz map above a threshold \citep{Yan2019}. Inspecting the Planck map by eye, a threshold of 0.006 K$_{cmb}$ was used. The idea behind the method is that the presence of the Musca cloud should give rise to a jump in reddening at a certain distance. The plot of the reddening as a function of the distance shows indication of such a jump at a distance of 140 pc, suggesting the presence of the Musca cloud. However, it can be noted that there already is reddening for a few stars before 140 pc. This generally starts at a distance of 90-100 pc and might indicate the presence of some gas that is more nearby than the Musca filament/cloud. Note that there generally is significant uncertainty about the reddening calculated in the GAIA catalogue \citep{Andrae2018}, such that this reddening effect before 140 pc might appear larger than it actually is. Fig. \ref{distMusca} shows that the nearby stars ($<$ 140 pc) with high reddening are located all over the map. This could imply that this is not the result of nearby gas concentrated in a small region, but that such nearby gas might be present over a larger area which could be related to the Chamaeleon-Musca complex. The same approach was used for Cha I, Cha II and Cha III, suggesting a distance of 180-190 pc, see Fig. \ref{distancesCham}, which is consistent with the results from \cite{Zucker2019}.

\section{The $^{13}$CO(3-2)/$^{13}$CO(2-1) uncertainty}
\label{app: densityModelling}
\begin{figure}
\includegraphics[width=\hsize]{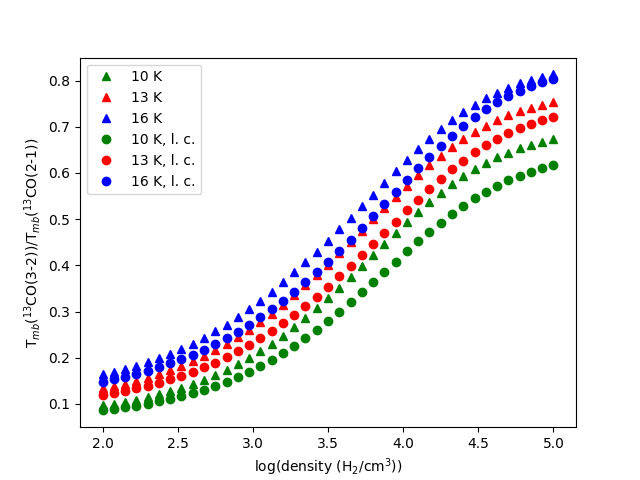}
\includegraphics[width=\hsize]{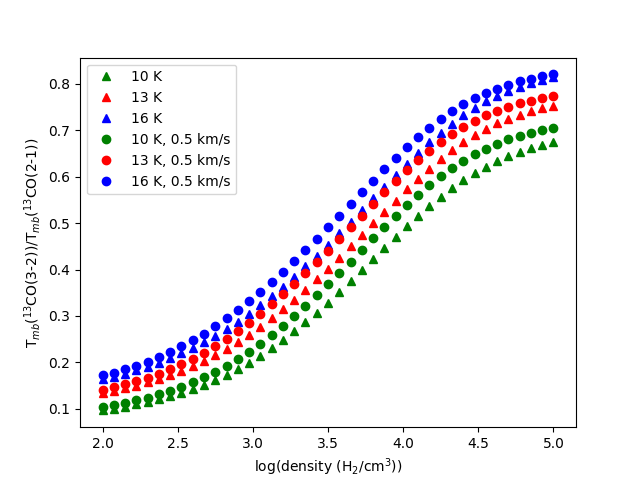}
\caption{\textbf{Top}: Impact of a lower column density, namely N$_{H_{2}}$ = 4$\cdot$10$^{21}$ cm$^{-2}$ (indicated with l. c.) instead of N$_{H_{2}}$ = 6$\cdot$10$^{21}$ cm$^{-2}$, on the $^{13}$CO(3-2)/$^{13}$CO(2-1) ratio as a function the density. \textbf{Bottom}: The impact of a lower FWHM (0.5 km s$^{-1}$ instead of 0.7 km s$^{-1}$) on the $^{13}$CO(3-2)/$^{13}$CO(2-1) ratios as a function of the density. For both the width and the column density, their impact is plotted for 3 different temperatures: 10 K, 13 K and 16 K. This shows that the impact of these two variables is smaller than the uncertainty related to the temperature.}
\label{errorDensity}
\end{figure}

To estimate the average density along the line of sight in the Musca filament and make an evolution versus distance from the filament crest, we have studied the ratio of $^{13}$CO(3-2)/$^{13}$CO(2-1). Here, we check the dependence of the density estimate on the width of the line and the column density of the line.\\% We will also show in more detail how the density as a function of the radius was constructed.\\
To study the impact of a change in width, the same setup as in section \ref{sec: density} was run, but this time with a FWHM of 0.5 km s$^{-1}$ (which is the minimal FWHM over the entire map). To study the impact of the column density, the setup from section \ref{sec: density} was retaken but this time with N$_{H_{2}}$ = 4$\cdot$10$^{21}$ cm$^{-2}$ (or 33\% less than the maximal column density used in section \ref{sec: density}). The impact of these changes are plotted in Fig. \ref{errorDensity}. When one has a fixed $^{13}$CO(3-2)/$^{13}$CO(2-1) ratio, for a lower column density this would result in a slightly higher density of the gas. The opposite is true for the FWHM: when the FWHM is lower for a fixed $^{13}$CO(3-2)/$^{13}$CO(2-1) ratio, the density for this ratio will be slightly lower than for a higher FWHM. When comparing the variation of the ratio as a function of density for a different FWHM or column density with the variation related to a different temperature in Fig. \ref{errorDensity}, it is found that the variation related to the temperature is larger than the variation related to the FWHM or column density. Furthermore, the FWHM and column density values used in section \ref{sec: density} are both relatively high for the strands located around the filament such that for both parameters a lower value might be more accurate. But the above analysis indicates that the effect of lowering the FWHM and column density at the same time might roughly cancel out due to their opposite impact on the ratio. The estimated densities in the strands might thus be reasonable as well. This is confirmed when using the obtained estimates for the density in the strands and filament from the above model to characterise the gas in these components. The intensity of the observed lines towards Musca are reproduced for temperatures and column densities that can be expected from \textit{\textit{Herschel}}. This provides further confidence that the density estimates might be relatively good.\\
%{\bf Furthermore, in Tab. \ref{brightnessRADEX} the effects of varying abundance ratios and column densities on the predicted line brightness are also presented. From these values it is observed that abundance and column density variations up to 50\% have a relatively small effect on the observed brightness compared to the temperature and density.}

\end{appendix}

\end{document}